# SPECTRALLY SHARP NEAR-FIELD THERMAL EMISSION: REVEALING SOME DISAGREEMENTS BETWEEN A CASIMIR-POLDER SENSOR AND PREDICTIONS FROM FAR-FIELD EMITTANCE

**J.C. de Aquino Carvalho [1, *], I. Maurin [1], P. Chaves de Souza Segundo [1, #], A. Laliotis [1], D. de Sousa Meneses [2], D. Bloch [1]**

[1] *Laboratoire de Physique des Lasers, UMR 7538 du CNRS, Université Sorbonne Paris Nord, 99 av. JB Clément, 93430 Villetaneuse, France*

[2] *CNRS, CEMHTI UPR 3079, Université d'Orléans, F-45071 Orléans, France*

[*] *Now at Departamento de Física, Universidade Federal de Pernambuco, Recife, PE 50670-901, Brazil,*
[#] *Permanent address: Centro de Educação e Saúde, Universidade Federal de Campina Grande, 581750-000 Cuité, PB, Brazil*

daniel.bloch@univ-paris13.fr

*Near-field thermal emission largely exceed blackbody radiation, owing to spectrally sharp emission in surface polaritons. We turn Casimir-Polder interaction between Cs(7P$_{1/2}$) and a sapphire interface, into a sensor sharply filtering, at 24.687 THz, the near-field sapphire emission at ~ 24.5 THz. Temperature evolution of sapphire mode is demonstrated. The Cs sensor, sensitive to both dispersion and dissipation, suggests the polariton to be red-shifted and sharper, as compared, up to 1100 K, to predictions from far-field sapphire emission, affected by birefringence and multiple resonances.*

The universality of the Blackbody Radiation (BBR) [1] led to energy quantization for matter and photons, founding the Quantum era. BBR applies in the Far-Field (FF) only, with respect to a wavelength $\sim hc/k_B T$ (50 µm for 300 K). At shorter distances [2-5], the contribution of evanescent fields, whose spectral features are material-dependent, dramatically enhances thermal exchanges; **similarly, the BBR limit can be beaten for small emitters [6,7]. This Near-Field (NF) regi**me opens fascinating applications [4-15], such as smart thermal infrared source (frequency, directivity, high-speed modulation) [16-17], coherent thermal emission controlled by surface engineering (gratings or metafabrication [18-22]), harvesting of thermal energy or refrigeration [23,24], thermal imaging [25-28], thermal rectification [29-31]. Ultimately, for a "close contact" transfer at atomic scale, distinguishing "radiation" and "conduction" becomes questionable [32-35] because the evanescent field originates in thermally-populated *surface polaritons* [36,37], *i.e.* hybrid particles coupling optical excitation and phonons originating in lattice excitation, hence involving a *phonon* transfer.

Spectral selectivity associated to the narrow-band NF thermal fluctuations [5], which affects both dissipative and dispersive effects, is essential for many applications. However, quantitative evaluations are few, often limited to very specific systems designed for nano-thermal engineering [19,22]. Indeed, monitoring thermal exchanges between macroscopic bodies, or stabilizing a temperature difference, is challenging at nanometric distances [10,14]. NF scanning microscopy allows a local probing of the thermal field even at high-T [40] but the nano-probe itself perturbs the polariton spectrum [5, 28, 41].

The thermal electromagnetic fluctuations are constrained by boundary conditions imposed by surface shape and material properties, so that spatial variations of NF thermal exchanges depend primarily on surface modes, and Maxwell propagation. Hence, for sake of simplicity, we concentrate here on the *complex* surface response for an interface between a planar homogeneous hot material and vacuum, $S(\omega)$:

$$S(\omega) = [\varepsilon(\omega) - 1]/[\varepsilon(\omega) + 1] \qquad (1)$$

with $\varepsilon(\omega)$ the macroscopic (relative) permittivity. As a major issue, further discussed, surface resonances occur for $\varepsilon(\omega) \rightarrow -1$, far away from bulk resonances [8, 42-45].

NF thermal emission is now encompassed in Fluctuating Electrodynamics [46], which also covers Casimir *dispersion* forces [47,48] describing the macroscopic quantum attraction at short distances between two polarizable neutral bodies. Casimir interaction, originally described for T = 0, results from the boundary conditions for vacuum fluctuations. Precision Casimir experiments now test fundamental predictions (standard model or hypothetical non-Newtonian gravity [49-51]), but demand $T \neq 0$ corrections and a realistic material description [52-54]. The analogous "Casimir-Polder" (C-P) interaction [55,56] between a neutral surface and a distant atom, benefiting of the intrinsic accuracy of Atomic Physics, has received various investigations in the very short distance regime –*i.e.* dipole-dipole van der Waals (vW) interaction. Among achievements, resonant couplings with surface modes [43,57,58], and temperature-dependence, were notably demonstrated [59,60].

Here, we analyze the narrow NF emission of sapphire ($Al_2O_3$), a technologically important material, up to T~ 1100 K, by comparing two radically different methods:

(i) Through a quasi-coincidence [43,57] between the sapphire surface emission and the Cs $7P_{1/2} \rightarrow 6D_{3/2}$ transition ($\omega_{at}/2\pi = 24.687$ THz, *i.e.* ~ 12.15 μm or ~ 823 $cm^{-1}$ [43,57,58,61] – see inset of fig.1), the C-P interaction on Cs($7P_{1/2}$) at short distance (~ 100 nm) narrowly filters the polariton thermal emission at $\omega_p/2\pi$ ~ 24.5 THz (~ 1200 K) [43,58].

(ii) Alternately, we evaluate surface polariton properties, *i.e.* $S(\omega)$ from eq.(1), with $\varepsilon(\omega)$ extrapolated from the broadband FF thermal emission.

In the short distance vW limit, the C-P interaction varies as - $C_3.z^{-3}$ ($z$: atom-surface distance [43,45,59-61]. The interaction, dominated by infrared transition couplings (see *Supp.*

*Mat.* in [62]), strongly varies when the thermal emission of the surface polariton (at $\omega_p$) nearly coincides with an *absorption* transition (at $\omega_{at}$ —dipole strength $D_{at}$) [59,60,62], following:

$$C_3(T) - C_3(T=0) = - 2\ D_{at}{}^2.\Re e[S(\omega_{at})].[\exp(\hbar\omega_{at}/k_BT) -1]^{-1} + o(T) \quad (2)$$

In eq.(2), the dominant right-hand-side term depends on the thermal Bose-Einstein factor, and on the surface resonant response (*underline*, *i.e.* dispersion) filtered at $\omega_{at}$; the $o(T)$ term, residual for $k_BT \geq \hbar\omega_{at}$, includes the nonresonant dipole couplings, and a $D_{at}{}^2$ contribution remaining insensitive to the surface resonance.

To measure $C_3$, Frequency-Modulated Selective Reflection spectroscopy (FM-SR) at a vapor interface is a robust method, probing a typical depth $\sim \lambda_{opt}/2\pi$ [57,58,60,61,86,87]. The measurement results from an optimized fitting of the experimental spectrum, on the basis of a single-parameter family of dimensionless FM-SR spectra [62]. We perform FM-SR on the second resonance doublet of Cs $6S_{1/2} \rightarrow \{7P_{1/2},7P_{3/2}\}$ (respectively $\lambda_{opt} = 459$ nm, $\lambda_{opt} = 456$ nm) [61]. As a spectroscopic method, FM-SR yields a differential coefficient $\Delta C_3 = [(C_3 (|7P>) - C_3 (|6S_{1/2}>)]$, here dominated by $C_3(|7P>)$, *i.e.* $C_3(|7P_{1/2}>)$ or $C_3(|7P_{3/2}>)$ [61]. $Cs(7P_{3/2})$ is primarily investigated for comparison purposes, as only coupled to the remote wing of the sapphire polariton –through $7P_{3/2} \rightarrow \{6D_{3/2},6D_{5/2}\}$ transitions, respectively at 19.259 THz and 20.544 THz, see inset of fig.1 and [62].

Figure 1 synthesizes our $C_3 (|7P>,T)$ measurements. For $Cs(7P_{1/2})$, $C_3(T)$ nearly triples from 500 K to 1100 K, while it gently decreases for $Cs(7P_{3/2})$. This growth (owing to $\Re e[S(\omega_{at})] < 0$) evidences the resonant coupling between $Cs(7P_{1/2})$ and sapphire thermal emission. It slows down at high-T: this signs a temperature modification of the surface resonance, *i.e.* a T-dependence for $\Re e[S(\omega_{at},T)]$; otherwise, $C_3(T)$ should follow the statistical factor of eq.(2), governing the dashed and dashed-dotted lines (fig.1) derived with eq.(1) for a constant sapphire permittivity -as measured at ambient temperature only [88-89] (see also [90]). Figure 1 also shows predictions (dotted and thick lines) with a T-dependent sapphire permittivity ([45,91,92] and discussion in [62]): as a trend, the agreement with our NF atomic sensor determinations for $C_3(T)$ is better, but remains relative, even for the thick line derived from our own measurements on sapphire windows from the very same origin than the one of the Cs vapor cell (identical super-polishing and annealing, see [60,62]). Ironically, when considering this careful evaluation of $S(\omega_{at},T)$ –detailed below–, the agreement is slightly worse than for predictions (dotted line) considering a permittivity $\varepsilon(\omega,T)$ [91, [62]] evaluated from measurements limited to the mid-infrared transparency region, on sapphire samples of an unknown origin, and modeled with an additional extrinsic resonance.

From FF thermal emission, that we have recorded with a state-of-the-art spectrometer [93], the complex-valued $\varepsilon(\omega,T)$, yielding $S(\omega,T)$, is extracted by fitting [45,62] a modified Lorentz dielectric function model including Gaussian multiphonon contributions *via* self-energy functions (wavenumber-dependent dampings). Sapphire birefringence adds a touch of complexity: the window at Cs vapor interface is $c_\perp$ (cut perpendicular to the *c*-axis) to ensure the cylindrical symmetry of C-P interaction, yielding an effective permittivity $\varepsilon_{eff} = (\varepsilon_o.\varepsilon_e)^{1/2}$ [43, 94] ($\varepsilon_o$ and $\varepsilon_e$ respectively the ordinary and extraordinary axes permittivity). The number of intrinsic modes is imposed by crystallography, depending on orientation (4 for ordinary, 2 for extraordinary [43,45,88]). Figure 2 exemplifies the measured and calculated emissivity sapphire spectra (fig. 2a-2b) for T~ 500 K and T ~ 1000 K (more temperatures in [62]). No salient singularities on emissivity (figs 2a-2b) help to locate resonances for $\varepsilon(\omega)$ (fig 2c -2d), or $S(\omega)$ (fig. 2e).

Along with thermal enhancement of C-P dispersion forces involving *virtual* atom-surface exchanges, surface thermal emission should induce *real* resonant energy absorption $7P_{1/2} \rightarrow 6D_{3/2}$, with the same $z^{-3}$ spatial dependence [42,59,95]. The reverse $6D_{3/2} \rightarrow 7P_{1/2}$ transfer, analyzed as a Förster-like *quenching* of atomic excitation induced by an absorbing surface mode, had been demonstrated [96], nevertheless without quantitative measurements.

We have unsuccessfully attempted to detect directly, as in [96], this Cs(7P$_{1/2}$) *absorption* of thermal energy, through an induced Cs(6D$_{3/2}$) population. Rather, we have extended the use of *linear* FM-SR, from the evaluation of the *dispersive* part of the surface interaction, to quantitative information on the *dissipative* exchanges, checking the relative amplitude of the doublet components 6S$_{1/2}$→{7P$_{1/2}$, 7P$_{3/2}$}. No NF thermal energy transfer is expected for Cs(7P$_{3/2}$), while thermal transfer shortens the Cs(7P$_{1/2}$) lifetime, specifically implying a spatially-dependent optical width, from γ$_∞$ in free-space to γ$(z)$ = γ$_∞$ + ΔΓ(T). $z^{-3}$, with:

$$\Delta\Gamma(T) = 4\, D_{at}^2 . \Im m[S(\omega_{at})].[\exp(\hbar\omega_{at}/k_B T) - 1]^{-1} \qquad (3)$$

This lifetime shortening of Cs(7P$_{1/2}$), governed by $\Im m[S(\omega_{at})]$, follows the same statistical factor as eq.(2) [62], and the same $z^{-3}$ dependence. The additional spatial-dependence γ$(z)$ mostly reduces amplitude of FM-SR spectra, complexifying moderately the lineshape family with an additional parameter (see.[62]). Inserting our estimates (fig.2e and [62]) into eq. 3 yields ΔΓ(T=1000 K) ~ 40 kHz.µm$^3$: the amplitude of the 6S$_{1/2}$-7P$_{1/2}$ FM-SR signal should become, at high-T, about twice smaller than for the unaffected 6S$_{1/2}$-7P$_{3/2}$ transition (see fig.3). Indeed, close to ω$_{at}$, $\Re e[S(\omega,T)]$ and $\Im m[S(\omega,T)]$ evolve nearly parallel above ~ 500 K, yielding $\Im m([S(\omega_{at},T)] / |\Re e[S(\omega_{at},T)]|$ ~ 1-2 [62]. Actually, the experimental amplitude ratio (fig.3) agrees with T = 0 predictions, once standard geometrical ratios for fine and hyperfine components [58,61,97] are included. In view of our measurements accuracy, this implies $\Im m[S(\omega_{at})] / |\Re e S(\omega_{at})| ≤ 0.1$-$0.2$ .

The discrepancies between our NF measurements and predictions from FF emissivity fitting cannot be explained by uncertainties affecting temperature measurements, atom dipole couplings, or the $C_3$(T) evaluation from FM-SR spectra (see [62]). Rather, relatively to expectations from fig.2e, we suspect the polariton resonance to be sharper and red-shifted, because, close to a surface resonance, $S(\omega)$ is merely a complex Lorentzian [98], with $\Im m S(\omega)$ and $\Re e S(\omega)$ associated respectively to absorption and dispersion. Hence, $\Im[S(\omega_{at})]/|\Re S(\omega_{at})| << 1$ implies that $\omega_{at}$ falls away from the anomalous dispersion region, with the peak amplitude for $S(\omega)$ largely exceeding $\Re[S(\omega_{at})]$; $\Re[S(\omega_{at})] < 0$ confirms that $\omega_{at}$ lies in the blue wing of the surface resonance. This hypothetical conclusion appears not incompatible with previous observations benefiting of the same Cs/sapphire coincidence at lower temperatures [62]. A complementary test for sapphire could be offered with Rb($7P_{3/2}$), through the dual resonant couplings $7P_{3/2} \to \{6D_{3/2}, 6D_{5/2}\}$, respectively at 24.494 THz and 24.561 THz [58].

Finally, for Cs($7P_{3/2}$), $C_3(T)$ appears slightly above predictions (fig.1), here insensitive to the specific modeling of sapphire polariton. This contrasts with the very good agreement found previously for Cs($7D_{3/2}$) [60,62], on the same cell, for a very comparable situation. However, the main T-dependent coupling was at 27.7 THz, in the opposite (blue) wing of the polariton, away from the multiple resonances of sapphire polariton. Uncertainties on $S(\omega,T)$ could be more severe for the red wing of $\omega_p$, where $\varepsilon_o(\omega)$ and $\varepsilon_e(\omega)$ differ considerably because of birefringence (see fig. 2c-d, major resonances respectively at ~13.5 THz and ~17 THz, *vs.* ~12 THz). In particular, the major $\varepsilon_e$ resonance falls partly outside the effective measurement range (~ 12-60 THz, see [62]), *i.e.* $\varepsilon_e(\omega)$ is extrapolated, rather than resulting from a genuine fitting. Additionally, the $\varepsilon_e$ resonance appears narrower than the two main resonances for $\varepsilon_o$ (fig. 2d), and this width impacts the remote wing of $\varepsilon_{eff}$, up to $\omega_{at}$.

To summarize, we demonstrate strong magnification of the CP interaction for Cs($7P_{1/2}$) through the resonant coupling to the NF thermally emitted surface polariton. The observed $C_3(T)$, not amenable to Bose-Einstein statistics, evidences a temperature evolution of surface resonance. The FM-SR linearity allows comparing the very different behaviors of the Cs(7P) doublet components: it reveals unique to provide simultaneously information related to dispersive and dissipative contributions of the surface emission. Predictions classically derived from the fitting of FF thermal emission at high-T for $\Im[S(\omega_{at})]$ disagree with FM-SR measurements, so that we suspect the actual surface resonance to be sharper than expected and slightly red-shifted. This discrepancy also appears with alternate descriptions of bulk sapphire properties: extrapolation of FF emission to the surface resonance is fragile, and here, birefringence and sapphire multiple resonances bring extra-difficulties.

Our results tackle the issue, of a general interest, of accuracy when evaluating surface mode resonances and NF spectral emission. Aside from the large shift between surface and bulk resonances, extrapolating $\varepsilon(\omega,T)$ from broadband FF emissivity (or reflectance) [45, 88] is intrinsically delicate as the complex-valued permittivity has to be derived from a real-valued spectral measurement. This is permitted by the Kramers-Kronig relationship but requires in the principle to know the entire spectrum. Classically, on a limited spectral range (5-25 µm in our case), a "guessed" analytical shape is optimized for permittivity, leading to a variety of sets of fitting parameters. The resulting ambiguity induces only negligible effects for $\varepsilon(\omega)$ and $S(\omega)$ in our modeling [62]. However, all models imply hard-to-test assumptions to describe the very remote wings of $\varepsilon(\omega)$ resonance, essential for evaluating $S(\omega_p)$. This would make valuable a direct measurement of the complex refractive index, in the range of the expected surface resonance. Ellipsometry is appropriate for this purpose, but remains underdeveloped in the infrared range and at high-T. The few corresponding measurements rely mostly on "spectroscopic ellipsometry", based upon an *a priori* description of the permittivity (see *e.g.* [89,99]).

Even if we have restricted ourselves to the extreme NF regime, and to thermal exchanges where one of the "materials" (Cs) is a narrow filter ideally known, uncertainties issues for surface resonances remain extremely relevant for longer distances, or for nanostructured interfaces—including exchanges between 2-dimensional layered materials [100]. Indeed, unexpected surface responses were often observed [25], evoking complex mechanisms, including non local response [101], strain [88], hot electrons within colder phonon vibrations [102], local glass structure instead of a crystalline one [103], along with chemical residuals [104,105]. Here, the consistency of fitting FM-SR spectra with a $z^{-3}$ potential is confirmed, in a regime dominated by the surface thermal emission: this makes very credible that Cs atoms effectively interact with a neutral surface; oppositely, when deriving $\varepsilon(\omega)$ from reflectance measurement, ideal "bulk" homogeneity and local planarity are implicit, ignoring issues of atomistic surface reorganization [106] or low-dimensionality [107].

Evaluating accurately NF thermal emission, including surface resonances, should benefit to various prospects:

(i) In Casimir research, oversimplified models describing material properties (*e.g.* Drude *vs.* plasma [108]) lead to known discrepancies: literature and tabulated data (*e.g.* [109]) should be scrutinized critically. For C-P interaction, NF material properties will govern the fine temperature tuning, turning vW attraction into repulsion [see *e.g.* 45].

(ii) Converting thermal surface emission into a quantized excitation (here, the unobserved $7P_{1/2}{\rightarrow}6D_{3/2}$ absorption) would open exciting possibilities [110,111], among which selective thermal NF population of molecular vibration modes (*e.g.* SiC surface modes at 28.5 THz coincide with $NH_3$ or $SF_6$ vibrations): this "surface thermal pumping" would ultimately generate specific (non "Boltzmann") internal energy distributions.

(iii) Birefringence, which notably affects sapphire, but SiC too, is shown here to induce dramatic changes of thermal infrared emission bands with orientation [112]. Engineering NF thermal conductivity, isolation, or rectification by orientating micro-bodies from the same material [10], appears exciting: presently, predicting the orientation-dependence of surface resonances remains challenging [113].

**Acknowledgements**

We thank D. Sarkisyan and his group (Ashtarak, Armenia) for the fabrication of the high-T Cs vapor cell. The stay of J.C. de A. C. was supported by the Brazilian grant *Ciências sem Fronteiras*. Various steps of this work were enriched with frequent discussions with J. R. Rios Leite, H. Failache, M. Chevrollier, M. Oriá, T. Passerat de Silans, P. Echegut. We also acknowledge discussions on near-field thermal emission and measurements with J-J. Greffet and Y. de Wilde, We also acknowledge the advice of J-M. Raimond for editing the manuscript. This research triggered support by the MITI interdisciplinary program of CNRS ("xNF High-T emission")

**Figure Captions**

Fig. 1 Experimental evaluations of $C_3$(T) for Cs(7P) derived from FM-SR spectroscopy on the $6S_{1/2} \rightarrow \{7P_{1/2}, 7P_{3/2}\}$ doublet (see inset and relevant dipole couplings), and predictions. For the $6S_{1/2} \rightarrow 7P_{1/2}$ transition, the 4 hyperfine components are plotted separately (▲: $4 \rightarrow 4$, ▲: $4 \rightarrow 3$; ▼: $3 \rightarrow 3$, ▼: $3 \rightarrow 4$); for the $6S_{1/2} \rightarrow 7P_{3/2}$ transition, ■ and ■ are respectively for the $3 \rightarrow \{2,3,4\}$ and $4 \rightarrow \{3,4,5\}$ manifolds; Cs reservoir temperature is 180°C. The dashed and dashed-dotted lines are predictions based upon sapphire permittivity at ~ 300 K (respectively [88] and [89]), the dotted ones considers a T-dependent permittivity [91]; the thick ones use our sapphire measurements.

Fig. 2 : Sapphire spectrral response at ~500 K (blue), and ~1000 K (red) —for exact temperatures and more details, see [62]. Experimental emissivity spectrum (reliable above 12 THz, see [62]), and analytical fittings (black): (a) $c_\perp$ sapphire window, yielding $\varepsilon_o$; (b) $c_{//}$ sapphire window and parallel polarization, yielding $\varepsilon_e$. Relative permittivity $\varepsilon(\omega)$ extrapolated from analytical fits of emissivity: (c) $\varepsilon_o(\omega)$; (d) $\varepsilon_e(\omega)$; (e) Complex surface response $S(\omega)$ calculated for $\varepsilon_{eff}(\omega) = (\varepsilon_o.\varepsilon_e)^{1/2}$ (full line), along with ~ 500K responses $S_o(\omega)$ and $S_e(\omega)$ [from $\varepsilon_o(\omega)$ and $\varepsilon_e(\omega)$] (respectively dashed, and dotted line); Note the additional horizontal scale in cm$^{-1}$, and μm, provided for conveniency. The vertical markers show the dominant couplings: full line for Cs(7P$_{1/2}$), dotted lines for Cs(7P$_{3/2}$).

Fig. 3: Temperature-dependence of the FM-SR amplitude ratio [$6S_{1/2} \rightarrow 7P_{1/2}$ over $6S_{1/2} \rightarrow 7P_{3/2}$]; amplitudes are normalized to $C_3 = 0$, and to the geometrical factors (fine and hyperfine manifolds [61]). The discrete triangles are predictions evaluated from our $S(\omega,T)$ (fig. 2e and [62]), the thick line is evaluated for $S(\omega,T)$ derived from [91]. The light gray area around the thick line accounts for amplitude ratio dependence with the optical width (central part: $\gamma_\infty = 20$ MHz).

Figure 1

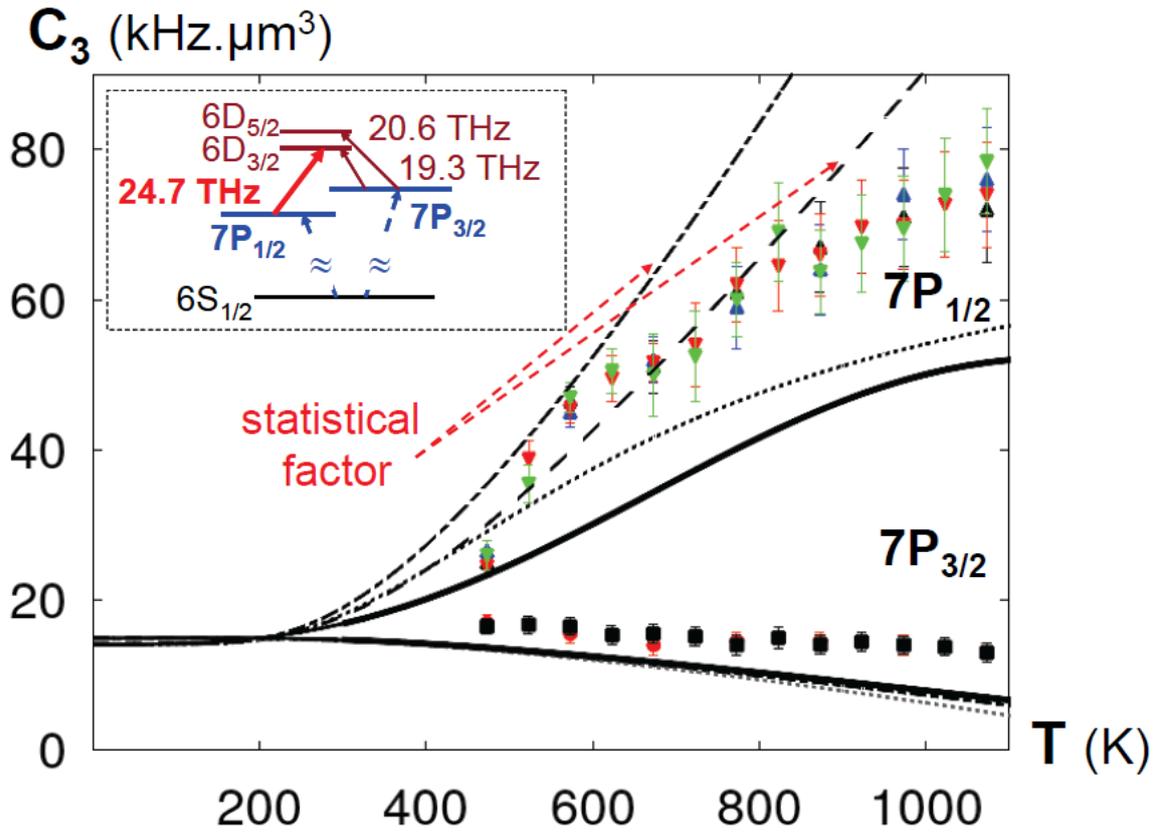

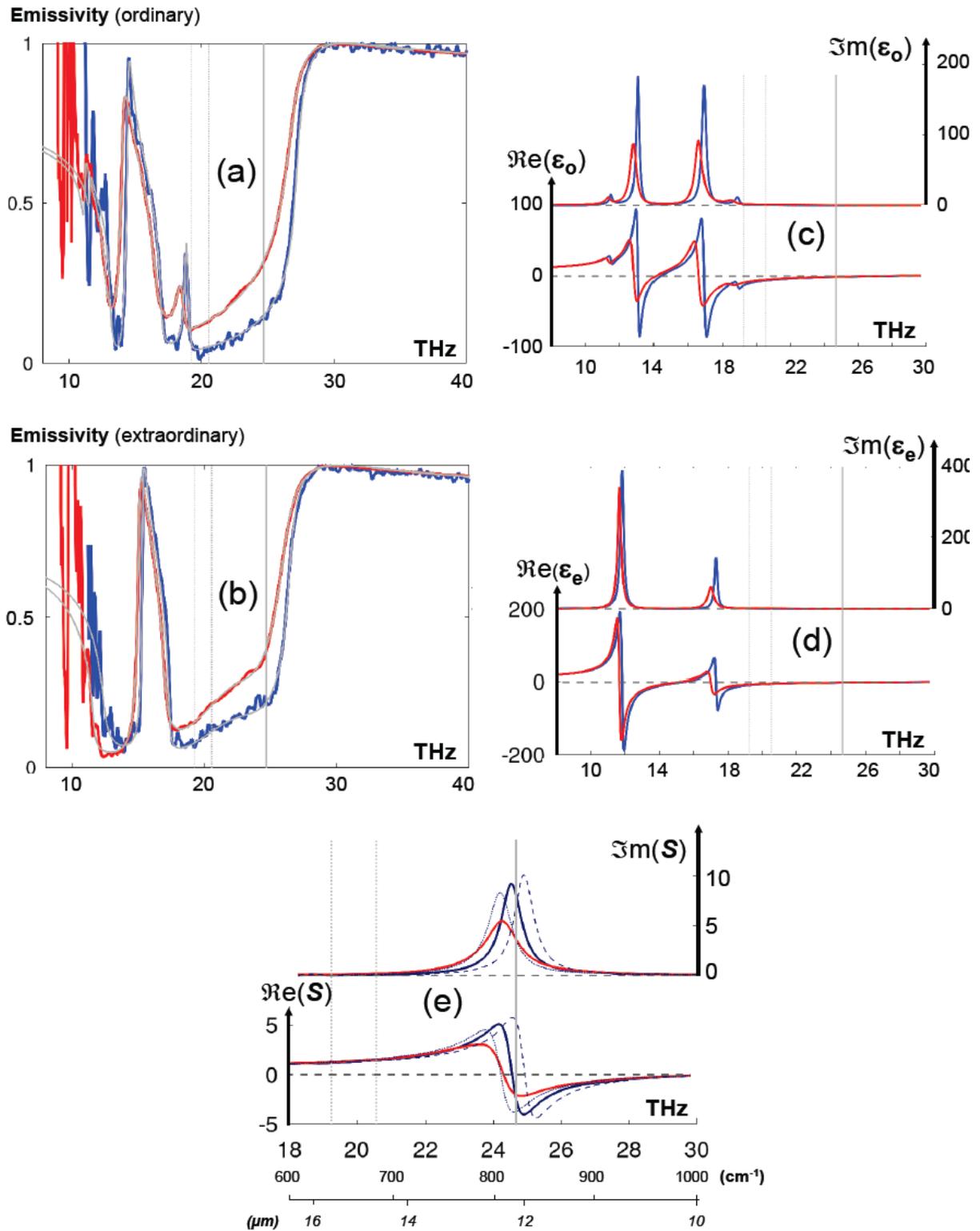

Figure 2

- 25 -

Figure 3

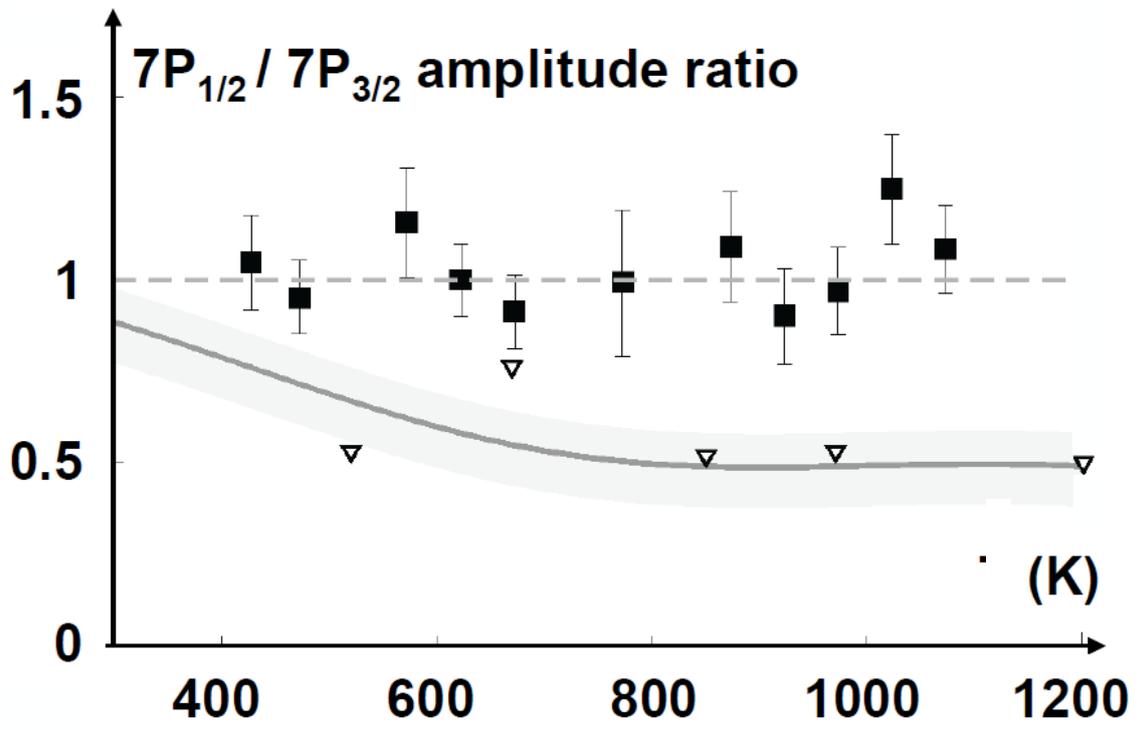

SUPPLEMENTARY MATERIAL for
**"Spectrally Sharp Near-Field Thermal Emission: revealing some disagreements between a Casimir-Polder sensor and predictions from Far-Field emittance"**

**J.C. deAquino Carvalho [1,\*], I. Maurin [1], P. Chaves de Souza Segundo [1,#], A. Laliotis [1], D. de Sousa Meneses [2], D. Bloch [1]**

[1] *Laboratoire de Physique des Lasers, UMR 7538 du CNRS,*
*Université Sorbonne Paris Nord, 99 av. JB Clément, 93430 Villetaneuse, France*
[2] *CNRS, CEMHTI UPR 3079, Université d'Orléans, F-45071 Orléans, France*
[\*] *Now at Departamento de Física, Universidade Federal de Pernambuco, Recife, PE 50670-901, Brazil,*
[#] *Permanent address: Universidade Federal de Paraíba – Campina Grande,*
*Centro de Educação e Saúde, Cuité, 58175-PB, Brazil*

This Supplementary Material addresses with more details various points raised in the primary *Letter*. It is organized in three sections: the first one is devoted to the Cs energy levels in the vicinity of a surface -notably hot sapphire- with energy shift and induced transfer rate, the second one is dedicated to the experimental evaluation of $C_3(T)$ (through spectroscopy), and the last one addresses the evaluation of the sapphire surface response from emissivity measurements.

## I. Atomic Physics and predictions for Surface Interaction
 1) Energy structure and couplings
 2) van der Waals regime of the Casimir-Polder interaction: $C_3$ predictions
  *a) General considerations*
  *b) Predictions concerning Cs(7P), Cs($6S_{1/2}$), accuracy of dipole transitions*
 3) Thermal energy transfer: prediction for Cs($7P_{1/2}$)
 4) Critical analysis of previous data and comparison with predictions

## II Experimental details regarding the spectroscopic measurements on Cs vapor
 1) General experimental set-up
 2) Atomic reference through saturated absorption
 3) Cs density in the multi-oven cell
 4) Principles for FM-SR spectroscopy
  *a) SR spectroscopy generalities*
  *b) Sub-Doppler SR and FM-SR*
  *c) Extrapolating the vW potential from FM-SR*
  *d) Data accumulation*
  *e) FM-SR interpreted with other (non vW) potentials*
 5) From FM-SR spectra to the extrapolation of vW measurements
  *a) Accuracy of the fittings and of the resulting $C_3$ estimates*
  *b) Comparison between FM-SR amplitudes on the $6P_{1/2} \rightarrow \{7P_{1/2} - 7P_{3/2}\}$ doublet*
  *c) Systematic uncertainties in FM-SR and how it may affect $C_3$ evaluation*

## III. From emissivity measurements to Temperature-dependent surface response of sapphire
 1) Sapphire samples and the spectrometer
 2) Sapphire birefringence
 3) From emissivity measurements to the dielectric permittivity $\varepsilon(\omega)$
 4) From permittivity $\varepsilon(\omega)$ to the surface response $S(\omega)$
 5) Uncertainty issues and comparison with literature
  *a) Experimental uncertainties affecting the emissivity spectrum, and the related permittivity*
  *b) Systematic attempts to modify the fitting values*
  *c) Other sources from literature*

## I. Atomic Physics and predictions for Surface Interaction

### 1) Energy structure and couplings

The energy structure of the $7P_{1/2}$ and $7P_{3/2}$ levels of Cs [1-3] is provided in fig. S-1. It shows the relevant hyperfine structures; the major dipole couplings appear with thicker lines.

The relative strength of hyperfine components is given in Table S-1 and the dipole couplings for the $7P_{1/2}$ and $7P_{3/2}$ levels are provided in the first columns of Table S-2, along with the recommended uncertainty when available [4].

| | | 7P$_{1/2}$ | | 7P$_{3/2}$ | | | |
|---|---|---|---|---|---|---|---|
| | | F' = 3 | F' = 4 | F' = 2 | F' = 3 | F' = 4 | F' = 5 |
| 6S$_{1/2}$ | F = 3 | 7/12 | 7/4 | 20/24 | 21/24 | 15/24 | forbidden |
| | F = 4 | 7/4 | 5/4 | forbidden | 7/24 | 21/24 | 44/24 |

Table S-1 : The relative absorption from $6S_{1/2}$ (F) to $7P_{J'}$ (F') (J' = 1/2, 3/2), with the included degeneracy factor of the ground state (respectively 7/16 and 9/16 for the F=3 and F=4 components of $6S_{1/2}$) (from [3]).

### 2) van der Waals regime of the Casimir-Polder interaction: $C_3$ predictions

#### a) General considerations

In the near-field van der Waals (vW) regime, the Casimir Polder (C-P) interaction for a |i> level, characterized by its $-C_3^i z^{-3}$ dependence, requires the knowledge of the strength of all relevant dipole couplings $D_{ij}$ for |i>→|j> dipole-allowed transition [7] along with the respective image coefficients $r(\omega_{ij})$ of the dielectric surface [8], with $C_3^i \propto \Sigma_j r(\omega_{ij}).D_{ij}^2$. These coefficients are temperature-dependent [9].

Table S-2 provides all the relevant information for the $C_3(T)$ prediction, for Cs($7P_{1/2}$) and Cs($7P_{3/2}$), at an interface with an ideal reflector [i.e. $r(\omega_{ij}) = 1$ whatever is $\omega_{ij}$ and for any temperature], and with sapphire. In this case, $r(\omega_i,T)$ depends on the sapphire permittivity $\varepsilon(\omega)$ spectrum –whose temperature-dependent evaluation is detailed in section III—, and on the photon occupation number Bose-Einstein temperature statistics through:

$$\bar{n}(\omega_{ij},T) = [\exp(\hbar\omega_{ij}/k_BT) - 1]^{-1} \qquad (S.1)$$

for $\omega_{ij} > 0$ ( |i> → |j> transition in absorption)

and $\bar{n}(\omega_{ij},T) = \exp(-\hbar\omega_{ij}/k_BT)/[\exp(-\hbar\omega_{ij}/k_BT) - 1] = 1 + \bar{n}(|\omega_{ij}|,T) \qquad (S.2)$

for $\omega_{ij} < 0$ (|i>→|j> transition in emission)

Indeed, one has:

$$C_3^i = (4\pi\varepsilon_0)^{-1}(1/12h)(2J_i + 1)^{-1}.\Sigma_n |<n_iJ_i|\mathbf{D}|n_jJ_j>|^2. r(\omega_{ij}, T) \qquad (S.3)$$

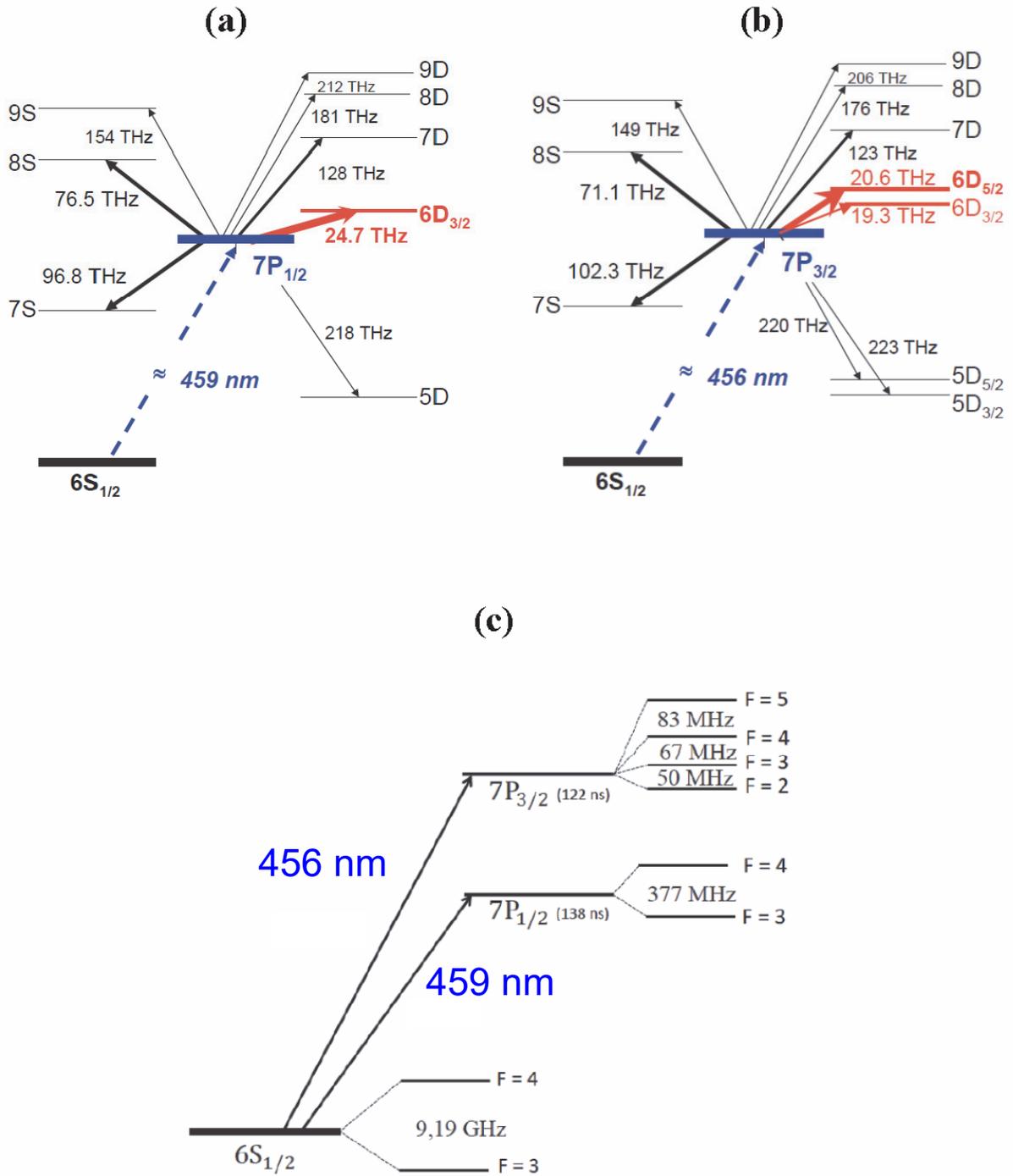

Fig. S-1: Atomic energy levels and couplings for Cs second resonance $6S_{1/2} \rightarrow \{7P_{1/2}, 7P_{3/2}\}$: (a) main dipole couplings for Cs ($7P_{1/2}$); (b) main dipole couplings for Cs ($7P_{3/2}$); (c) Details on the optical transitions (second resonance line of Cs) allowing to reach the 7P level, with a blow-up of the hyperfine structure; 456 nm stands for 657.9 THz, and 459 nm for 652.5 THz.

**7P$_{1/2}$**

| coupled level $\lvert i \rangle$ | $\omega_{at} = \omega_{7P} - \omega_i$ THz | K | Dipole value from [4] or *extrapolated from [5] a.u | ±(%) | $C_3$ ideal reflector kHz.μm³ | $C_3$ sapphire 300 K $\Re eS(\omega_{at})$ | $C_3(\omega_{at})$ kHz.μm³ | 520 K $\Re eS(\omega_{at})$ | $C_3(\omega_{at})$ kHz.μm³ | 670 K $\Re eS(\omega_{at})$ | $C_3(\omega_{at})$ kHz.μm³ | 850 K $\Re eS(\omega_{at})$ | $C_3(\omega_{at})$ kHz.μm³ | 970 K $\Re eS(\omega_{at})$ | $C_3(\omega_{at})$ kHz.μm³ |
|---|---|---|---|---|---|---|---|---|---|---|---|---|---|---|---|
| 6S$_{1/2}$ | - 652.5 | | 0.45* | | 0.01 | | 0.0048 | | 0.0048 | | 0.0046 | | 0.0049 | | 0.0048 |
| 7S$_{1/2}$ | - 96.83 | 4646 | 10.31 | 0.39 | 4.32 | | 1.69 | | 1.62 | | 1.44 | | 1.52 | | 1.44 |
| 8S$_{1/2}$ | 76.50 | 3647 | 9.31 | 0.25 | 3.52 | | 2.10 | | 2.17 | | 2.25 | | 2.35 | | 2.42 |
| 9S$_{1/2}$ | 154.25 | | 1.96 | 0.91 | 0.13 | | 0.073 | | 0.074 | | 0.076 | | 0.077 | | 0.078 |
| 5D$_{3/2}$ | - 217.8 | | 2.3 | (+18) | 0.17 | | 0.078 | | 0.077 | | 0.072 | | 0.076 | | 0.074 |
| 6D$_{3/2}$ | 24.687 | 1184 | 17.99 | 0.25 | 13.55 | -5.23 | 11.97 | -2.67 | 20.06 | -2.68 | 28.74 | -2.53 | 39.14 | -2.2 | 43.35 |
| 7D$_{3/2}$ | 128.39 | 6162 | 6.6 | 3.38 | 1.77 | | 1.01 | | 1.03 | | 1.05 | | 1.08 | | 1.10 |
| 8D$_{3/2}$ | 181.25 | | 3.15* | | 0.4 | | 0.22 | | 0.23 | | 0.23 | | 0.23 | | 0.23 |
| 9D$_{3/2}$ | 211.75 | | 1.98* | | 0.16 | | 0.09 | | 0.09 | | 0.09 | | 0.09 | | 0.09 |
| **Total** | | | | | **23.63** | | **17.32** | | **25.45** | | **34.04** | | **44.66** | | **48.88** |

**7P$_{3/2}$**

| coupled level $\lvert i \rangle$ | $\omega_{at} = \omega_{7P} - \omega_i$ THz | K | Dipole value from [4] or *extrapolated from [5] **extrapolated from [6] a.u | ±(%) | $C_3$ ideal reflector kHz.μm³ | $C_3$ sapphire 300 K $\Re eS(\omega_{at})$ | $C_3(\omega_{at})$ kHz.μm³ | 520 K $\Re eS(\omega_{at})$ | $C_3(\omega_{at})$ kHz.μm³ | 670 K $\Re eS(\omega_{at})$ | $C_3(\omega_{at})$ kHz.μm³ | 850 K $\Re eS(\omega_{at})$ | $C_3(\omega_{at})$ kHz.μm³ | 970 K $\Re eS(\omega_{at})$ | $C_3(\omega_{at})$ kHz.μm³ |
|---|---|---|---|---|---|---|---|---|---|---|---|---|---|---|---|
| 6S$_{1/2}$ | -657.9 | | 1.21* | | 0.01 | | 0.005 | | 0.005 | | 0.005 | | 0.005 | | 0.005 |
| 7S$_{1/2}$ | -102.3 | | 14.32 | 0.43 | 4.17 | | 1.66 | | 1.60 | | 1.43 | | 1.51 | | 1.43 |
| 8S$_{1/2}$ | 71.07 | 3410 | 14.07 | 0.48 | 4.02 | | 2.41 | | 2.51 | | 2.60 | | 2.72 | | 2.81 |
| 9S$_{1/2}$ | 148.8 | | 2.67** | | 0.14 | | 0.078 | | 0.080 | | 0.081 | | 0.083 | | 0.085 |
| 5D$_{3/2}$ | -223.26 | | 0.9 | 21.12 | 0.01 | | 0.005 | | 0.005 | | 0.005 | | 0.004 | | 0.004 |
| 5D$_{5/2}$ | -220.33 | | 2.8 | 19.24 | 0.11 | | 0.05 | | 0.05 | | 0.05 | | 0.05 | | 0.05 |
| 6D$_{3/2}$ | 19.259 | 924 | 8.07 | 0.26 | 1.33 | 1.18 | 0.90 | 1.22 | 0.78 | 1.31 | 0.64 | 1.28 | 0.41 | 1.28 | 0.26 |
| 6D$_{5/2}$ | 20.544 | 986 | 24.35 | 0.24 | 12.04 | 1.44 | 7.95 | 1.48 | 6.10 | 1.51 | 4.22 | 1.50 | 1.23 | 1.55 | -0.58 |
| 7D$_{3/2}$ | 122.96 | | 3.3 | 3.19 | 0.22 | | 0.13 | | 0.13 | | 0.13 | | 0.14 | | 0.14 |
| 7D$_{5/2}$ | 123.59 | | 9.6 | 3.05 | 1.87 | | 1.07 | | 1.09 | | 1.11 | | 1.14 | | 1.16 |
| 8D$_{3/2}$ | 175.82 | | 1.48** | | 0.04 | | 0.022 | | 0.022 | | 0.023 | | 0.023 | | 0.024 |
| 8D$_{5/2}$ | 176.17 | | 4.42** | | 0.40 | | 0.22 | | 0.23 | | 0.23 | | 0.23 | | 0.24 |
| 9D$_{3/2}$ | 206.33 | | 0.92** | | 0.02 | | 0.011 | | 0.011 | | 0.011 | | 0.012 | | 0.012 |
| 9D$_{5/2}$ | 206.53 | | 2.76** | | 0.15 | | 0.082 | | 0.084 | | 0.085 | | 0.086 | | 0.087 |
| **Total** | | | | | **24.63** | | **14.59** | | **12.76** | | **10.62** | | **7.66** | | **5.73** |

Table S-2: Estimates for $C_3(T)$ for Cs(7P$_{1/2}$) and Cs(7P$_{3/2}$), with the contributions of the most important dipole couplings. The tables provide values for an ideal reflecting surface, and for sapphire at selected temperatures, taking into account our determinations for $S(\omega)$. We provide our estimated values of $\Re eS(\omega, T)$ for the dominant transitions (for temperatures averaged between the respective temperatures of *ordinary* and *extraordinary* sapphire measurements —see section III). The transition frequency is also provided in temperature units for the lowest energetic couplings, indicating the typical temperature for which an effect may occur.

with $n_i$ and $n_j$ respectively the principal quantum number of $|i>$ and $|j>$, $J_i$ the electronic angular momentum ($J = L+S$) of $|i>$ level, and **D** the dipole operator. Note that in eq. (S.3), $C_3^i$ is expressed in frequency [*i.e.* $(2\pi)$.Hz, and not in circular frequency, *i.e.* rad/s].

The temperature-dependent image coefficient $r(\omega_{ij}, T)$ for each contribution is the sum of two contributions, a resonant one $r_{res}$ and a (dispersive) nonresonant one $r_{NR}$ [9]:

$$r(\omega_{ij}, T) = r_{res}(\omega_{ij},T) + r_{NR}(\omega_{ij},T) \quad \text{(S.4)}$$

In eq. (S.4), the resonant contribution $r_{res}(\omega_{ij},T)$ is sensitive to the surface response at the frequency of the atomic coupling $\omega_{ij}$ through:

$$r_{res}(\omega_{ij},T) = -\omega_{ij}/|\omega_{ij}|. \, 2 \, \Re e[S(|\omega_{ij}|)]. \, \bar{n}(\omega_{ij},T) \quad \text{(S.5)}$$

with $S(\omega)=[\varepsilon(\omega)-1]/[\varepsilon(\omega)+1]$.

The non resonant (dispersive) contribution $r_{NR}(\omega_{ij},T)$ verifies $0 \leq r_{NR}(\omega_{ij},T) \leq 1$ for $T = 0$ [8]. It encompasses the whole spectrum of the material by a continuous summing at $T = 0$, becoming [9] at $T \neq 0$, a discrete sum over contributions at the Matsubara frequencies $\xi_p = p \, 2\pi k_B T/\hbar$ ($p$ integer):

$$r_{NR}(\omega_{ij},T) = (4k_B T/\hbar). \sum_{p=0}^{\infty}{}' \, \frac{\varepsilon(i\xi_p)-1}{\varepsilon(i\xi_p)+1} \, \frac{\omega_{ij}^2}{\omega_{ij}^2+\xi_p^2} \quad \text{(S.6)}$$

In eq. (S.6), $\Sigma'$ is a sum symbol with the first term taken only to half its value.

*b) Predictions concerning Cs(7P), Cs(6S$_{1/2}$), accuracy of dipole transitions*

For the $C_3(T)$ predictions, Atomic Physics data (*i.e.* tabulated data) are required for each $|i> \rightarrow |j>$ dipole coupling involved in the summing in eq. (S.3), as well as the $\bar{n}(\omega_{ij},T)$ value, and the knowledge of the image coefficient $r(\omega_{ij},T)$. In the extended tables (S-2), a column shows the contribution of each notable dipole coupling for an ideal reflector. Such "ideal reflector" contributions were already provided in [7]. Here, the present values use refined dipole evaluations. [4]. Also, the dependence of $C_3$ on the hyperfine component is negligible [7].

In table S-2, the indicated values for $r(\omega_{ij},T)$ are derived from eqs. (S-4)-(S-6) and from our sapphire measurements detailed in the section III. Table S-2 amply justifies the approximate expansion of eq. 2 (*Letter*) for Cs(7P$_{1/2}$), with the temperature dependence of $C_3(T)$ purely dominated by the 7P$_{1/2}$-6D$_{3/2}$ coupling, in a strong coincidence with the sapphire surface resonance. For Cs(7P$_{3/2}$), table S-2 shows that at least the couplings to the doublet {6D$_{3/2}$, 6D$_{5/2}$} need to be considered, and the overall temperature dependence is moderate.

In spectroscopic measurements, the FM-SR technique on an optical transition $|a> \rightarrow |b>$ is sensitive to the difference $C_3(|b>) - C_3(|a>)$. This makes it essential for our experiments to get the precise value of the $C_3$ coefficient for Cs (6S): close to a sapphire interface, one has $C_3(6S_{1/2}) =$ 1.3 kHz.$\mu$m$^3$ [*i.e.* $C_3(6S_{1/2}) << C_3(7P)$], with an entirely negligible temperature dependence [2].

Note also that here, with literature estimating the uncertainty in the theoretical dipole couplings, the global uncertainty for $C_3(T)$ is estimated to be <0.25%, notably because for Cs(7P$_{1/2}$), the dominant coupling to 6D$_{3/2}$ is known with an excellent accuracy. In other cases (see section I.4), revised estimates for dipole couplings can lead to revise $C_3$ estimates.

### 3) Thermal energy transfer prediction for Cs(7P$_{1/2}$)

In our study, the real energy transfer induced by a thermally excited surface corresponds to an atom *absorbing* the surface excitation on the 7P$_{1/2}$→6D$_{3/2}$ line, owing to the specific surface resonance of sapphire —in the same way, a coincidence between atom *emission* and surface excitation thermally stimulates the *decay* to the lower level [9].

The transfer rate $\Gamma_3(|i> \rightarrow |j>$, T) is the dissipative equivalent of the dispersive response associated to $C_3(T)$. It is governed, for a transition to an upper state, by :

$$\Gamma_3(|i> \rightarrow |j>,T) = (4\pi\varepsilon_0)^{-1} \, (1/3h).z^{-3}. \, (2J_i+1)^{-1} \, \bar{n} \, (\omega_{ij},T) \, |< n_iJ_i|\mathbf{D}|n_jJ_j>|^2. \, \Im m[S(\omega_{ij})] \qquad (S.7)$$

with $\omega_{ij}$ the frequency for the 7P$_{1/2}$→6D$_{3/2}$ transition, and $\bar{n}$ ($\omega_{ij}$,T) given by eq. (S.1).

Figure S-2 indicates, as a function of temperature, the transfer rate for Cs(7P$_{1/2}$) induced by the thermal excitation of sapphire. The essential information in fig. S-2 comes from $\Im m S(\omega_{ij}$,T), in addition to the dipole coupling already appearing in Table S-2 [see eqs. (S.7) and (S.5)], and to the number of thermal photons. In Fig. S-2, the considered values for $\Im m S(\omega_{ij}$,T) for sapphire are provided from our far-field measurements [see fig. 2e in *Letter*, or below in section III and fig. S-11] along with alternate models for sapphire [10,11].

This population transfer reduces the lifetime of the $|i>$ level. In the general theory of relaxation, the increase of optical width (*i.e.* coherence) should be only *half* of the increased population transfer rate. Here, for transitions starting from the ground state, the system is equivalent to a close (two-level) system —without population losses, despite the hyperfine pumping, with its very slow relaxation. Hence, the related increase in optical width $\gamma$, induced by the thermal transfer, adds to the width $\gamma_{opt}$ which also includes collision broadening, so that one has to consider $\gamma(z) = \gamma_{opt} + \Delta\Gamma(T).z^{-3}$, with $\Delta\Gamma(T) = \Gamma_3(|i> \rightarrow |j>,T).z^3$

### 4) Critical analysis of previous data and comparison with new predictions

In the first measurement [12] demonstrating a temperature dependence $C_3(T)$, performed on Cs(7D$_{3/2}$), we had attempted to evaluate the uncertainties on the dipole strengths. With the more recent prediction for the 7D$_{3/2}$→8P$_{1/2}$ dipole coupling [4], dominant in this last study, the prediction for $C_3(T)$ needs to be revised by $\sim +10$ kHz.$\mu$m$^3$: this still allows a very good agreement between theory and measurements, although not as excellent as initially indicated.

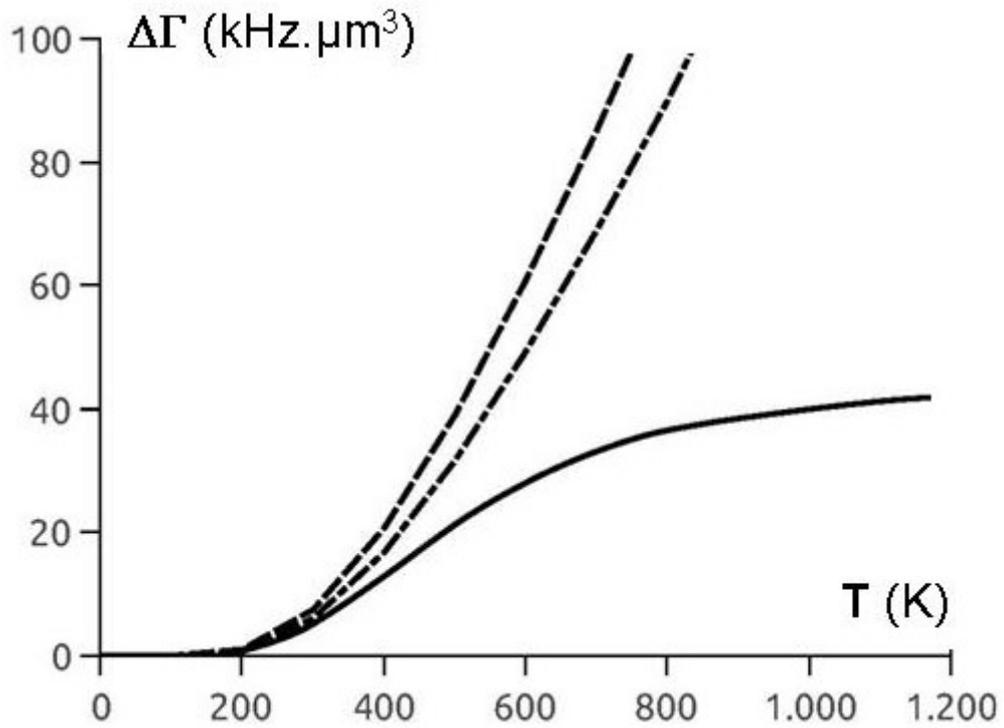

Fig. S-2 : Predictions for $\Delta\Gamma$ as a function of temperature T for Cs ($7P_{1/2}$) (*i.e.* transfer to $6D_{3/2}$). The predictions in full line are derived from our present estimates for $\Im m[S(\omega,T)]$, the dashed-dotted line and the dashed line, correspond respectively to values for $\Im m[S(\omega,T)]$ derived from sapphire permittivity as given in [10], and [11], as measured at room temperature only.

Another important comparison between experimental measurements of $C_3$ and predictions is for the situation where the sapphire resonance clearly leads to a vW-CP repulsion [13]. The sapphire predictions (- 100 kHz.$\mu$m$^3$) were found slightly below experimental findings (- 160 $\pm$ 40 kHz.$\mu$m$^3$), for evaluations at that time solely based on [10] (and [8]), *i.e.* sapphire properties evaluated only at room temperature, with an "optimal" sample (observations could vary with the sample). With our first temperature-dependent measurements on sapphire [14], evidencing temperature broadening and shift of the sapphire surface resonance, the strength of the repulsion should be lowered down to $\sim$ - 50 kHz.$\mu$m$^3$ if one retains the predictions for sapphire at T = 200°C. This strong revision can be however tempered as the actual temperature of the Cs window in these experiments was $\sim$ 130-150°C, and, because the nominal "200°C" temperature probably exceeded 200°C, owing to calibration issues for the temperature in the reflectivity spectrometer used in [14]. This discrepancy between the experimental $C_3$, and the revised sapphire predictions for this experiment does not contradict our suggestion (see *Letter*), that the actual sapphire resonance at high temperature could be sharper (and shifted), relatively to standard expectations as derived from far-field estimates for sapphire.

With the suggestion from *Letter* that the surface resonance is red-shifted, it may appear more difficult to understand the early demonstration [15] of a real energy transfer –through quenching– by resonant coupling between surface mode and atom emission. Actually, the demonstration, which has not provided any quantitative evaluation for $\Im m S(\omega_{at})$ -and has remained independent of $\Re e S(\omega_{at})$- was performed on a $c_{//}$ window. When the cylindrical symmetry typical of a $c_\perp$ window is lost, the quenching coupling can occur through $\Im m S_o(\omega_{at})$ or through $\Im m S_e(\omega_{at})$ as well. In this last case, the resonance is on the blue-side of $\omega_{at}$ (see *Letter*, fig.2 e, or section III fig; S-11): hence, the early experimental findings do not contradict the suggested possibility that the actual sapphire surface resonance is shifted to the red, relatively to expectations from the modeling of far-field properties.

**II Experimental details regarding the spectroscopic measurements on Cs vapor**

1) <u>General experimental set-up</u>

The set-up for the Frequency-Modulated Selective Reflection (FM-SR) experiments [2] is described in fig. S-3. Two different lasers are respectively used for the 459 nm and 456 nm transitions (External Cavity Diode Lasers ECDLs by Toptica, "DL100" for 459 nm, and "DL100 pro" for 456 nm). The frequency is tuned by a piezo actuator which acts on the cavity length. The two laser beams, optically isolated and reshaped after spatial filtering, follow the same optical paths, providing ideal conditions for a comparison between the two fine-structure transitions. FM is obtained by applying a ~ 1 kHz modulation to the piezo. In its principle, FM-SR spectroscopy requires only an elementary set-up, with a single incident beam sent at normal incidence (here ~ 2 mrad) to the cell window/vapor interface.

Frequency markers are provided by an auxiliary Saturated Absorption (SA) set-up, with a chopper on the pump beam, and by an ultra-stable 60 cm-long Fabry-Perot, used in a c/6L geometry (thanks to 2 spherical mirrors, with a 120 cm curvature radius). In the absence of applied FM, the SA spectra exhibit a truly sub-Doppler Lorentzian, with a width possibly as small as 1.5 MHz.

The hot Cs cell, described elsewhere [2,12], is made entirely of sapphire, but for the Cs reservoir, which is in glass and imposes limits to the local temperature ($\leq$ 200°C). It is ended by a superpolished annealed $c_\perp$ sapphire window ($c$-axis perpendicular to the window). The cell is heated-up by a system of three independent ovens, whose temperatures are computer-controlled. This ensures a very good stability (1-2°C) in operation, notably at the level of the Cs reservoir, minimizing the risks of amplitude fluctuations induced by variation of Cs pressure [16] –for successive recordings of identical spectra, or when comparing 459 nm and 456 nm spectra. The absolute temperature uncertainty is however much higher (~ 30-50 °C for the hot window), but this affects negligibly the overall temperature dependence, and our conclusions. Note also that thermal desorption probably helps to clean-up the window [12], minimizing surface defects [17].

2) <u>Atomic reference through saturated absorption</u>

To estimate the surface interaction in FM-SR, the SR spectrum is compared with a Doppler-free SA reference spectrum, obtained in a similar environment, but free of surface interaction. This volume reference is obtained in an auxiliary 1 cm-long Cs cell. All FM-SR spectra are recorded simultaneously with this auxiliary SA spectrum.

Because the Cs density increases very quickly with the Cs reservoir temperature [16], the signal amplitude can be increased on request. As a counterpart, pressure effects on the transition of interest for the vW measurement must be evaluated. Complementary SA experiments were performed on a 100 μm-thick heated Cs cell, short enough to allow a high density without an excessive absorption, while the SA signal on the 1-cm long cell was kept as a low-pressure reference. Our measurements (see [2] and fig. S-4) provide an evaluation of pressure effects for the $6P_{1/2} \rightarrow 7P_{1/2}$ transition, and confirm previous results [7] for the stronger $6P_{1/2} \rightarrow 7P_{3/2}$ transition. This pressure shift is only a small fraction of the broadening, nearly negligible with respect to the effect of the spatially-dependent surface-induced interaction. This remains true even with the high-temperature of the Cs cell window, when the higher thermal speed of the colliding atoms (see below) may increase the lineshape broadening and residual shift.

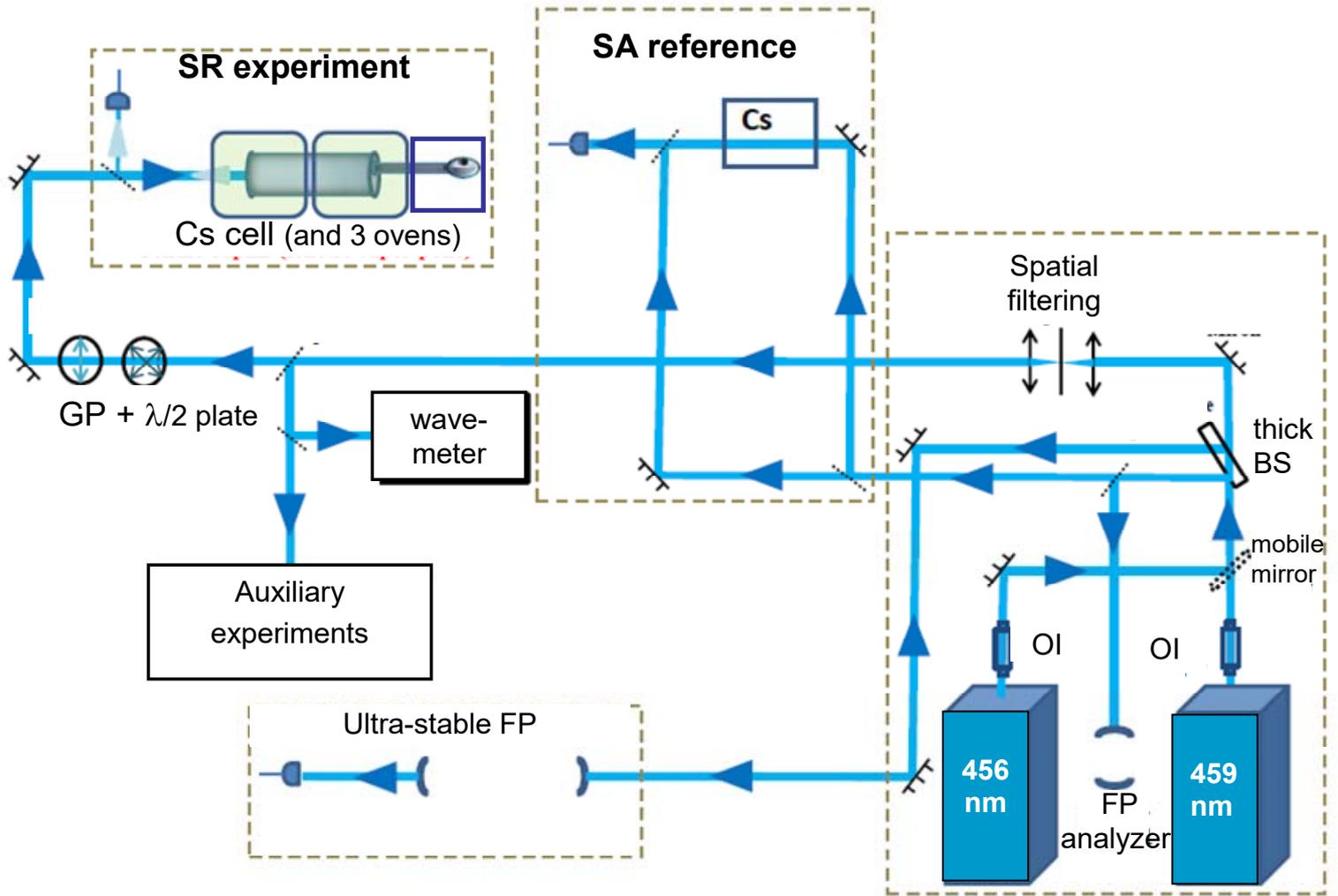

Fig. S-3 Schematics of the experimental set-up. The output of each laser goes first to an optical isolator (OI), while a mobile mirror allows selecting the 456 nm laser beam, or the 459 nm beam, while keeping the same alignment. The proper behavior o f  t h e  l a s e r  s c a n  i s  controlled through a wavemeter and an ultrastable Fabry-Perot (FP). The auxiliary beams are picked-up through a thick beam splitter (BS) at the laser output.  A FM is applied to each laser, and the SA reference and SR signals are both detected through a lock-in detection, demodulating the applied FM. For the SR experiment, the intensity is controlled through a λ/2 plate followed by a Glan prism (GP) polarizer. The auxiliary experiments can notably include auxiliary SA experiments for pressure broadening and shift.

### 3) Cs density in the multi-oven cell

Practically, the temperature dependence of interest in our experiments is associated to the temperature of the sapphire window, and of the surrounding oven, with which it is at thermal equilibrium [2,12]. The Cs vapor only provides a collection of individual Cs atoms, from which the $C_3$ value is assessed. The vapor itself is not in a macroscopic thermal equilibrium, owing to the high temperature of the sapphire window with respect to the lower temperature of the Cs reservoir. Rather, the situation corresponds to a *local* thermal equilibrium between the mean kinetic energy of the atoms of the vapor and the cell walls, as confirmed by independent experiments, performed through SR spectroscopy on the 894 nm resonance line. This is consistent with a pressure which remains constant all over the cell, at the expense of a locally varying Cs density, as defined by the local thermal equilibrium of an (ideal) gas, and hence governed at the reservoir by the local thermal equilibrium [12]. Note that for a previous and comparable work on this same Cs cell [12], the local equilibrium of the vapor had not to be considered, because the transition of interest was reached after a prior step of optical pumping, assumed to provide a quasi-thermalization. Also, in related experiments [18], the effect of blackbody radiation transfer in *volume*, which competes with the *near-field* thermal emission, was investigated. This volume effect, detectable for appropriate and higher temperatures and densities, remains fully negligible in our situation.

### 4) FM-SR spectroscopy: principle

#### a) SR spectroscopy generalities

FM-SR (linear) spectroscopy has been developed for a long time in our group (see [19] and refs. therein), in view of the evaluation of the vW surface interaction. Here, the novelty is in our interest in the amplitude comparison of the signals at 456 nm and 459 nm. Indeed, we can fully exploit the intrinsic linearity of the SR technique because the experiments are performed on one-photon transition from the ground-state, and not through a stepwise two-photon excitation, as in [12,13]. Also, the experiments are performed under a low intensity irradiation (typically $\leq$ 1 mW/mm²). This ensures linearity of the SR experiment (see fig. S-5) as the saturation intensity, indicated in [20] to be 0.5 mW/mm² for the 459 nm line, is here notably increased by the collision broadening, which dominates the natural width.

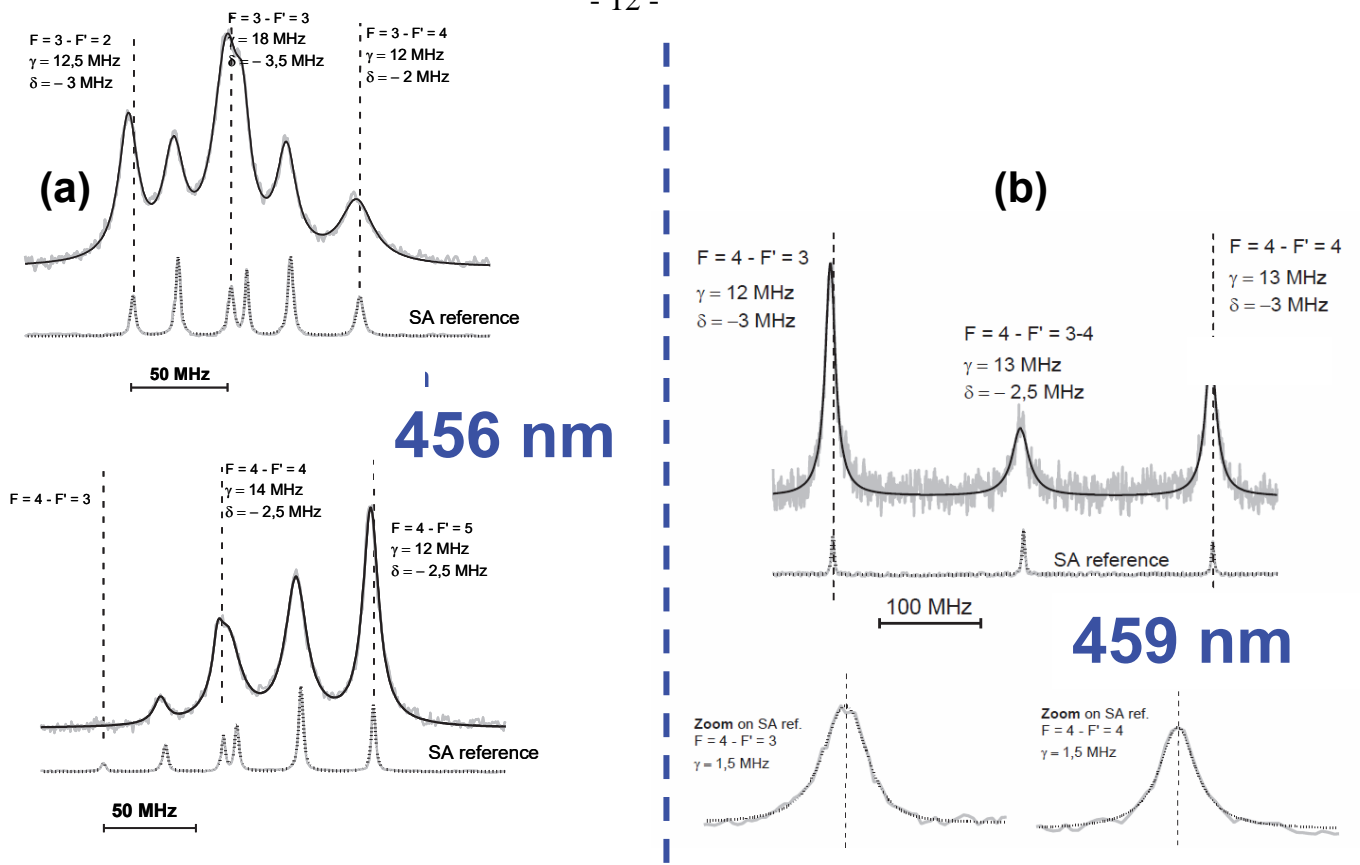

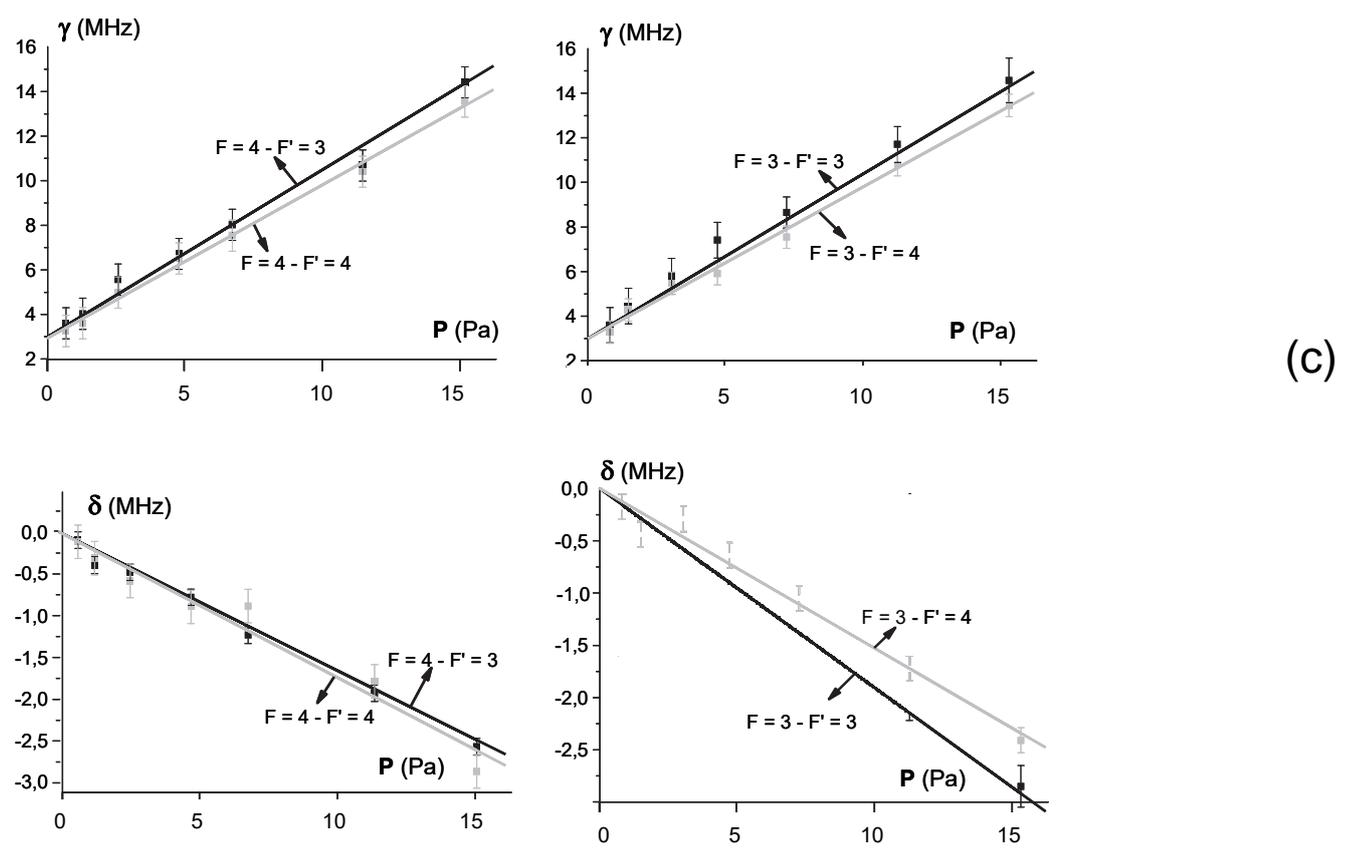

Fig. S-4: Pressure broadening ($\gamma$) and shift ($\delta$) of the 6S-7P doublet, as determined by comparing SA spectra in the 1 cm Cs cell at low temperature, and in a heated 100 µm-thick cell. The spectra are fitted with a Lorentzian for each individual component, *i.e.* h y p e r f i n e c o m p o n e n t o r S A Doppler crossover resonance): (a) typical spectra on the $6S_{1/2} \rightarrow 7P_{3/2}$ transition at 456 nm —some components partially overlap; (b) same as (a) for the $6S_{1/2}(F=4) \rightarrow 7P_{1/2}\{F'=3,4\}$ transition at 459 nm; (c) Synthesis of the pressure broadening and shift for all hyperfine components of the $6S_{1/2} \rightarrow 7P_{1/2}$ transition (for the corresponding results on the $6S_{1/2} \rightarrow 7P_{13/2}$ transition, see [7]).

*b) Sub-Doppler SR and FM-SR*

SR yields spectroscopic information on the gas layer at the interface, through a spatial integration of the local atomic response modulated by the complex factor exp(2i$kz$), with $k = 2\pi/\lambda_{opt}$ ($\lambda_{opt}$ the wavelength of the optical transition, here 456 nm or 459 nm). This means that SR is sensitive to regions where the atomic response varies spatially, *i.e.* regions close to the surface. The probed depth in SR is usually estimated to be $\sim \lambda_{opt}/2\pi$. However, if the surface interaction is strong enough, atoms can be totally off-resonance close to the surface, extending the region probed in resonant spectroscopy. Typical FM-SR resonances for Cs($7P_{1/2}$) span over several tens of MHz away (red-side) from the atomic resonance, so that the probed region is typically $\sim$100 nm for $C_3$ values in the 15-80 kHz.µm³ range.

Under normal incidence, the SR signal includes a specific sub-Doppler response, enhancing the response of atoms moving nearly parallel to the window. Moreover, with a small FM applied to the laser and the SR signal demodulated accordingly, the FM-SR signal is simply proportional to the frequency-derivative of the initial SR signal, and becomes Doppler-free [19,21,22] in the infinite width Doppler approximation. To optimize the FM-SR signal amplitude, the FM amplitude can be increased up to the limit of a lineshape broadening -or distortion- for the SR signal. In most cases, we have chosen a FM amplitude $\sim$ 3.5 MHz for the 459 nm laser, and $\sim$ 2.5 MHz for 456 nm laser [2]. This induces a negligible effect on the FM-SR lineshape, although this is sufficient for a broadening to appear on the narrower low-pressure SA reference spectrum.

*c) Extrapolating the vW potential from FM-SR*

For a spectrum with a single FM-SR resonance, as it occurs for the fully resolved hyperfine components for the $6S_{1/2}$-$7P_{1/2}$ line, $C_3$ is extracted by comparison with universal lineshapes, built with a single dimensionless parameter A characterizing the interaction strength [22]:

$$A = 2k^3 C_3/\gamma \qquad \text{(S.8)}$$

These universal lineshapes are made more realistic by summing-up the contribution of the various hyperfine components, notably when these components are not fully resolved [7,12,13]

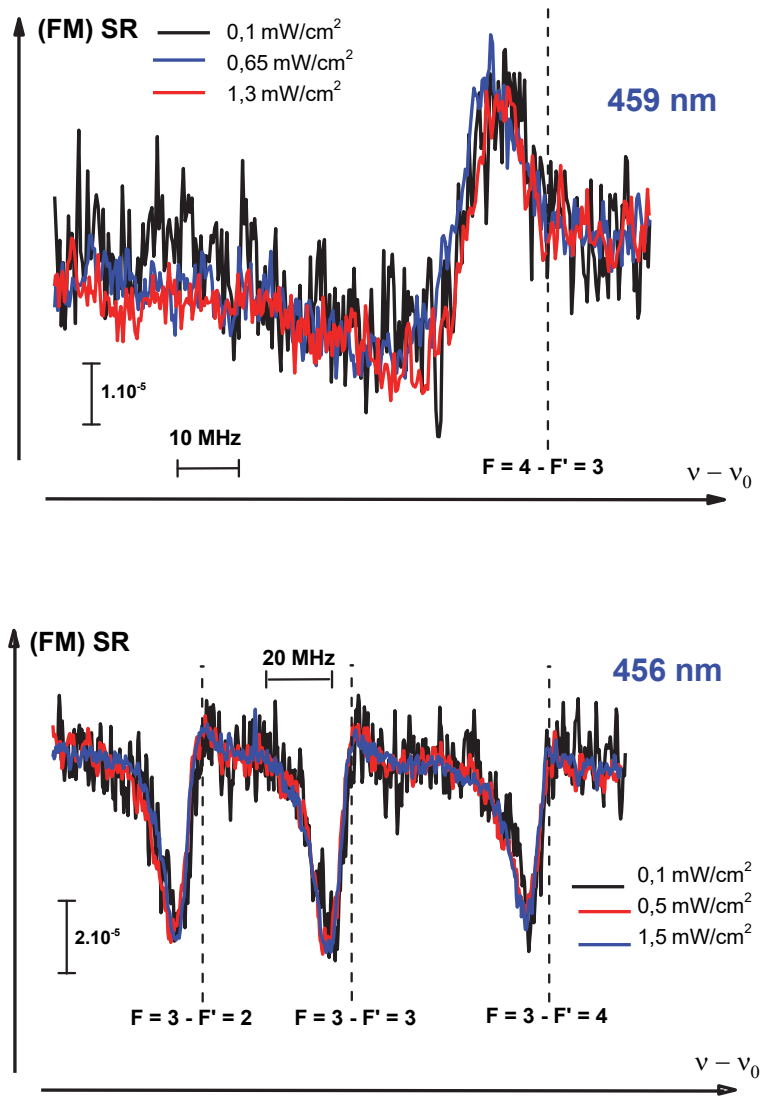

Fig. S-5: Experimental FM-SR spectra, as obtained for different irradiation intensities (as indicated) demonstrating a linear regime (*i.e.* SR lineshape not affected by irradiation intensity). The vertical axis indicates the relative change of reflected intensity (*i.e.* the FM amplitude is taken into account, to make the change of intensity independent of the FM amplitude): (a) $6S_{1/2}(F=4) \rightarrow 7P_{1/2}(F'=3)$ transition; (b) $6S_{1/2}(F=3) \rightarrow 7P_{3/2}\{F'=2,3,4\}$ transition.

*e.g.* at high Cs pressure on the $6S_{1/2} \rightarrow 7P_{3/2}$ transition. It is also possible to include a finite Doppler width correction, when the "infinite" Doppler width approximation does not hold [19,23]. Distinct values of the width ($\gamma$), shift ($\delta$) with respect to the volume resonance, amplitude ($\mathcal{C}_1$), and A coefficients, can be considered for individual components [23,24]. This reveals here useless for our hyperfine manifolds: identical widths and A values [7], with relative amplitudes governed by the geometrical factors of Table S-1, can be considered. The range of acceptable $C_3$ values is obtained by exploring discrete values of the "A" shape-factor, searching for each A value the least-square fitting theoretical lineshape (see fig. S-6, and [7,23]). The adjustable width $\gamma$, and frequency shift $\delta$ with respect to the SA reference (fig. S-7), account for pressure broadening and shift, but have to remain compatible with the independent evaluation discussed above (sub section II – 2).

A satisfactory fitting of a given FM-SR lineshape is not sufficient to demonstrate that the lineshape originates in a $z^{-3}$ potential. A consistency step is usually required, to ensure that the extrapolated surface interaction (*i.e.* $C_3$) does not vary with the vapor condition [7,23]. The significance of our FM-SR analyses has been demonstrated at length [7,13], and has relied on $C_3$ robustness under variations of the Cs pressure, and hence of the optical width $\gamma$ (through atom-atom collisions): for an atom-surface interaction independent of atom-atom collisions, pressure variations notable modifies the FM-SR lineshape, and A. This is why we have recorded numerous series of FM-SR spectra, to investigate the C-P effect at a given window temperature. For many of our investigations, FM-SR spectra were recorded by keeping the Cs reservoir at a given temperature, exploring a large range of surface temperature for the $c_\perp$ sapphire window. Series were performed successively with the Cs reservoir at 160°C, 180°C, 200°C (*i.e.* ~1.5, 5, 15 Pa), with complementary series with reservoir at 150°C and 170°C. Other complementary series were recorded alternately, at a specific window temperature and varying reservoir temperature. In all cases, the Cs density remains low, leading to a low signal level. Note additionnally that when increasing the window temperature –for a constant Cs reservoir temperature–, the SR amplitude clearly decreases: this agrees with the idea that pressure (but not the density) is constant all over the cell, despite the temperature gradient.

All hyperfine components, fully resolved for $7P_{1/2}$, have been shown to yield similar spectra. Most often, we have repeated systematically these independent measurements for $C_3$ [and hence for $\Re e S(\omega_{at})$], on all hyperfine components [2]. The slight variations between hyperfine components -see *Letter* fig.1- appear to be purely statistical, without any systematic effects with the window temperature.

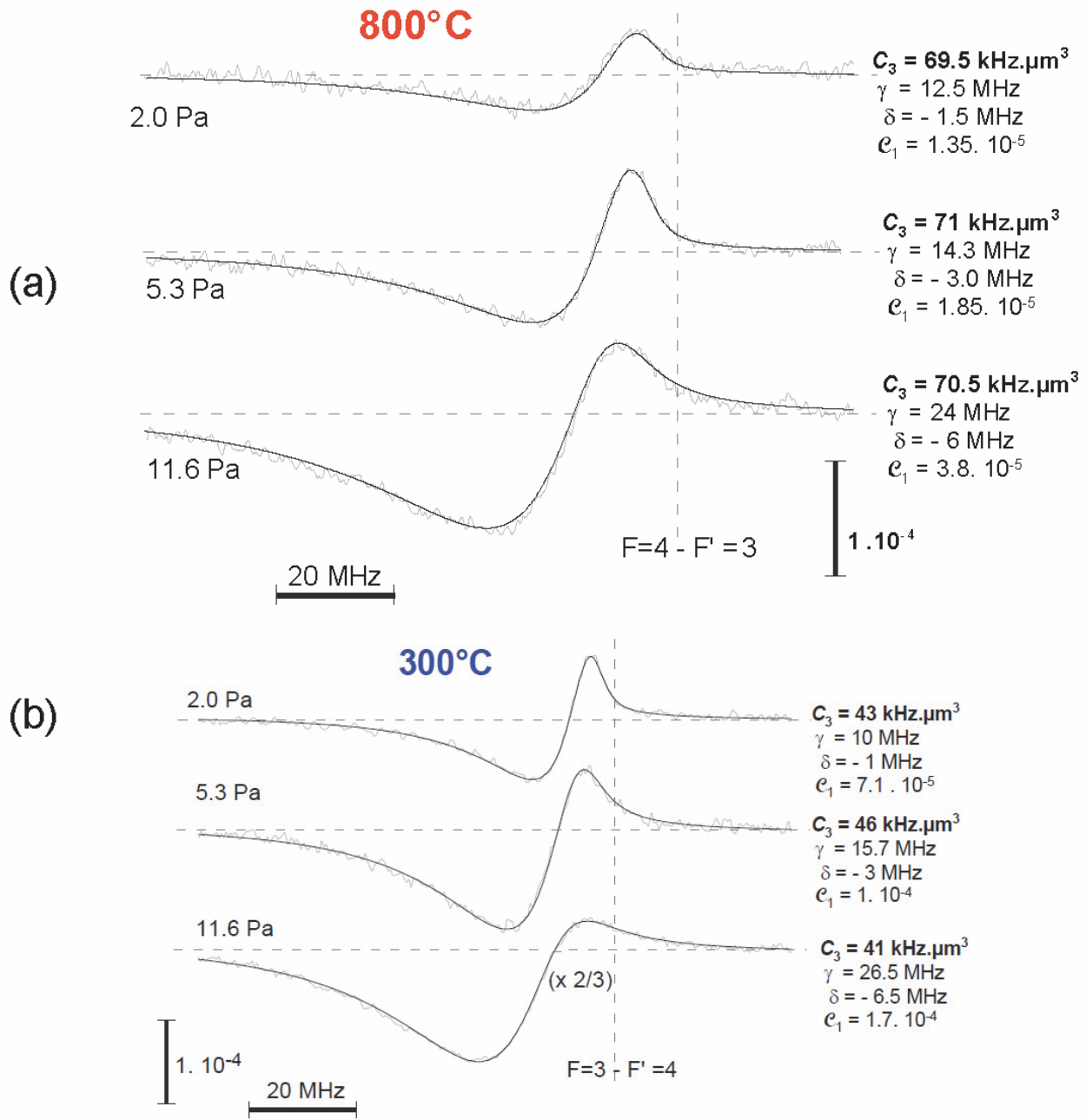

Fig. S-6: Examples of fully resolved hyperfine component of the FM SR spectra on the $6S_{1/2} \rightarrow 7P_{1/2}$ transition, for (a) T=800°C and F=4→F'=3; (b) T=300°C and F=3→F'=4. One observes a notable broadening with Cs pressure, as determined by the Cs reservoir temperature. The vertical dashed line provides the center of the SA reference spectrum (transition in volume). The optimal fitting lineshapes allow to extract simultaneously the values for C-P coefficient $C_3(7P_{1/2})$, width $\gamma$, shift $\delta$ relatively to the SA reference, and normalized amplitude $\mathcal{C}_1$. All hyperfine components exhibit a similar behavior.

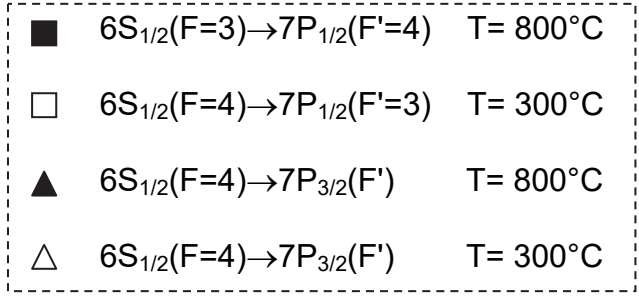

(a)

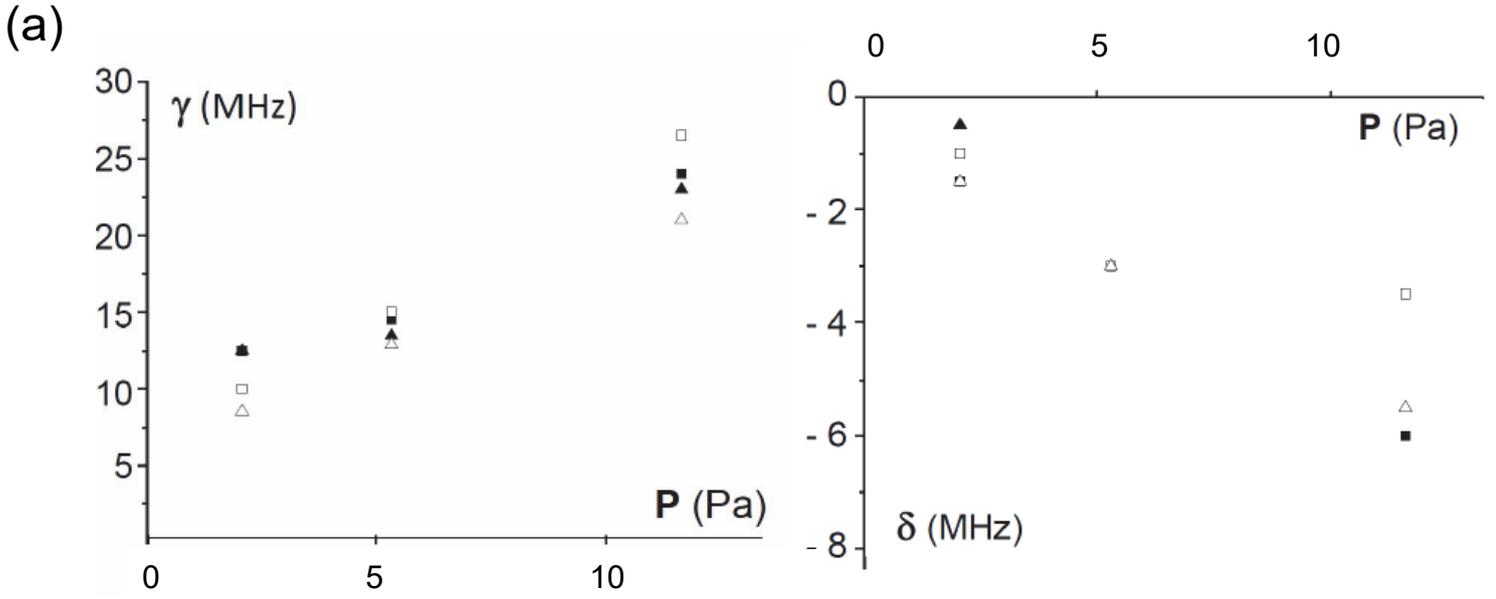

(b)

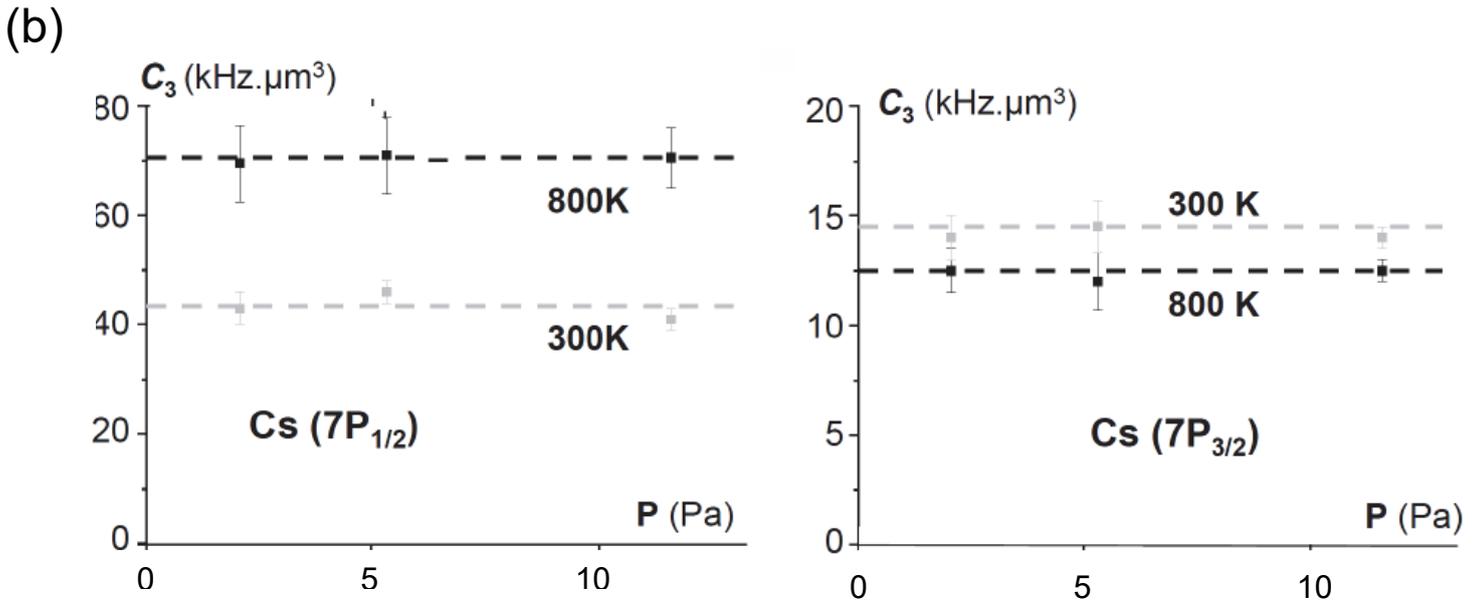

Fig. S-7: From the fitting of spectra such as shown in fig. S-6, one estimates: (a) the values for $\gamma$ and $\delta$ as a function of Cs pressure and (b) the $C_3$ values. The observed linear variations with pressure of $\gamma$ and $\delta$ are in agreement with the values shown in fig S-5, while the $C_3$ values are found to be, as expected, independent of the Cs pressure and only depending on the temperature at the interface [$7P_{1/2}$ (left) and $7P_{3/2}$ (right)].

*d) Data accumulation*

As a general challenge, the most significant spectra should be free of atom-atom collision effects, implying a low atomic density and low signal levels, and hence the need for data accumulation. The laser sources do not ensure a perfectly linear frequency scan, because of long-term drifts, precluding too slow scans of the laser frequency. Typical frequency scans are recorded in 2-3 minutes, and repeatedly (*e.g.* 10 times). To compensate for drifts and nonlinearity of the scans in the data accumulation, the frequency axis is rescaled, through the SA spectrum and its multiple hyperfine components and crossover resonances [2]. In addition, for the wings of the atomic references, the monitoring of the transmission through the long ultra-stable Fabry-Perot (free-spectral range: 83.33 MHz) enables an extra-recalibration of the frequency scan.

*e) FM-SR interpreted with other (non vW) potentials*

Theoretical FM-SR lineshapes can also be calculated [22] for a variety of surface potentials, not obeying the $-C_3z^{-3}$ dependence [25-27]. A fitting method, similar to the one mentioned above, can be applied for comparison between experimental spectra and such theoretical lineshapes. For instance, following [28], FM-SR lineshapes for $z^{-2}$ and $z^{-4}$ potentials have been calculated [26,27]. These potentials can be respectively found for the interaction with a charged surface, or with the far-field regime of the Casimir-Polder interaction for a ground-state atom. Satisfactory fittings are occasionally obtained [26], usually at the expense of an unjustified frequency shift $\delta$, with inconsistencies often appearing when the atomic density (*i.e.* $\gamma$) is varied on a large range.

Fluorescence into evanescent wave channels [29,30], moderately shortening Cs(7P) lifetime, may locally affect $\gamma$; this induces an apparent increase for $C_3$, which remains negligible as compared to the observed residual broadening ($\gamma \geq 8$ MHz, vs. 1.6 MHz natural width), independently of FM. Rather, as discussed in subsection I. 3, the surface thermal emission can modify the transition rates with the same $z^{-3}$ dependence found for the C-P potential. This justifies the eq. (3) in *Letter* (and section I.3). Universal FM-SR lineshapes can hence be calculated by introducing a second dimensionless parameter B [2,25] associated to this $z^{-3}$ dependence for the optical linewidth:

$$B = 2\,k^3\Delta\Gamma/\gamma \qquad (S.9)$$

Various couples of parameters (A, B) can yield very similar lineshapes. Sensible differences are nevertheless found in the width, *i.e.* fitting a given spectrum defined by a set of dimensionless parameters ($A_1$, $B_1$), with another lineshaped defined by a set ($A_2$, $B_2$), imposes to change the frequency scale, *i.e.* $\gamma_1$ into $\gamma_2$, with $\gamma_2 \neq \gamma_1$. Importantly, the amplitudes too are modified; the surface-specific extra-decay channel tends to reduce the signal amplitude. This appears in fig. S-8, showing fittings, with values of A relevant for our experiments (A $\leq$ 20), and B = 0 or B $\neq$ 0: the fitted amplitude $c_1$ increases with B, to compensate for the reduction in the amplitude of the

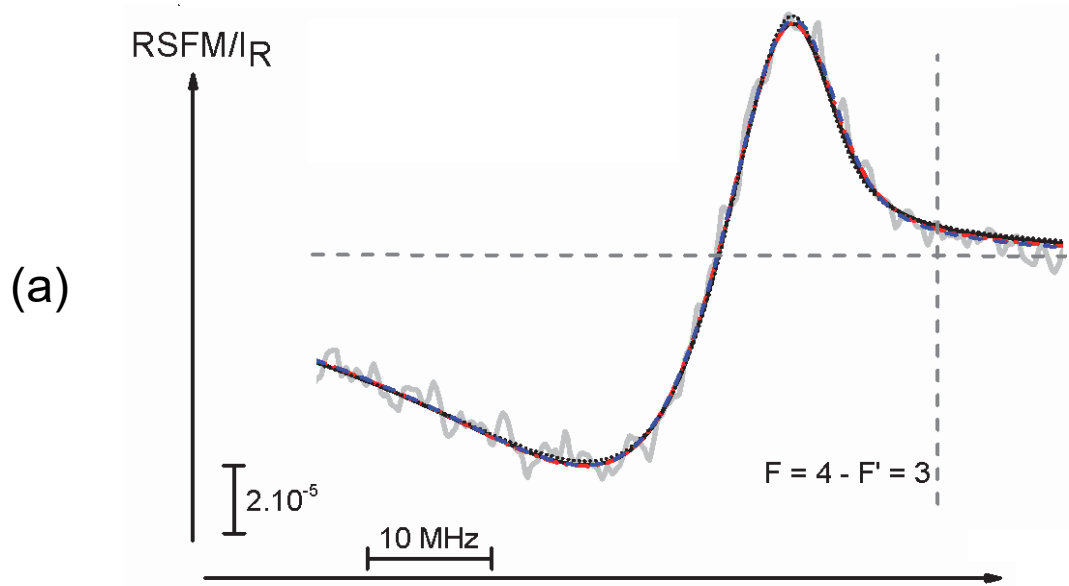

**A = 25.5, B = 0**, γ = 14.3 MHz, δ = -3 MHz $C_3$ = 71 kHz.μm³, $\mathcal{C}_1$= 2.10⁻⁵

**A = 25.5, B = 2**, γ = 14 MHz, δ = -3 MHz $C_3$ = 70 kHz.μm³, $\mathcal{C}_1$= 1.9.10⁻⁵

**A = 25.5, B = 8**, γ = 12.8 MHz, δ = -2.7 MHz $C_3$ = 64 kHz.μm³, $\mathcal{C}_1$= 3.10⁻⁵

**A = 25.5, B = 16**, γ = 11 MHz, δ = -2.5 MHz $C_3$ = 55 kHz.μm³, $\mathcal{C}_1$= 4.6.10⁻⁵

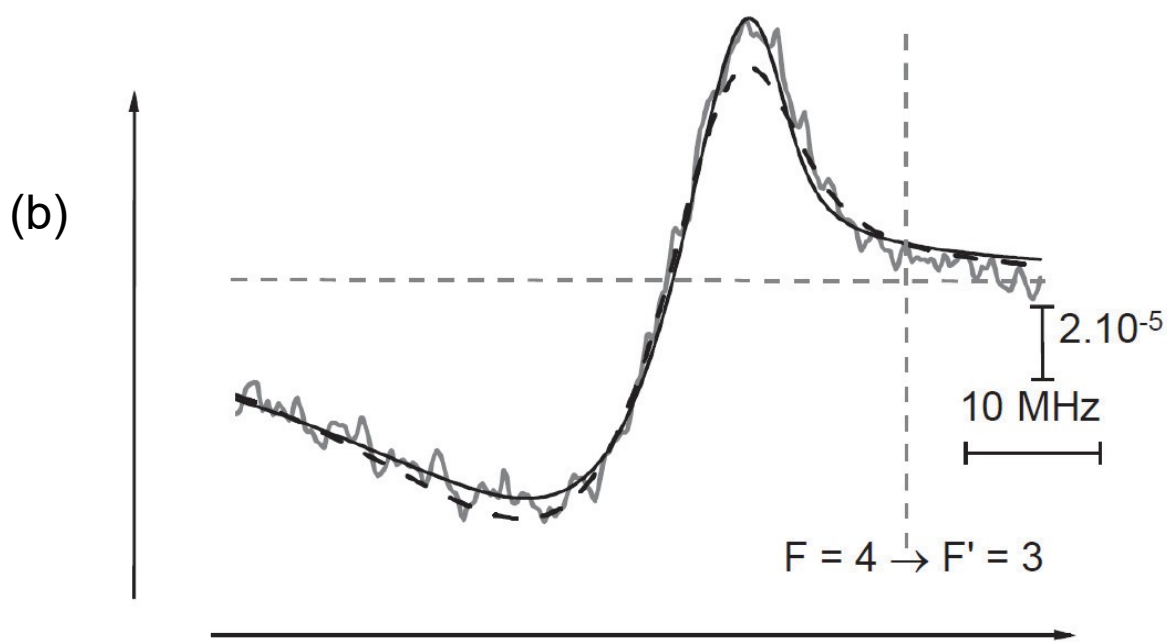

—— **A = 20, B =40**, γ = 8.9 MHz, δ = -3 MHz, $C_3$ = 35 kHz.μm³, $\mathcal{C}_1$= 8.9.10⁻⁵

- - - **A = 20, B = 0**, γ = 16 MHz, δ = -3.5 MHz, $C_3$ = 62 kHz.μm³, $\mathcal{C}_1$= 1.85.10⁻⁵

Fig S-8: Attempts to fit a given FM-SR signal —here, $6S_{1/2} \to 7P_{3/2}$ at T = 800°C at 5.3 Pa, already shown in fig. S-6— with a 2 dimensionless parameters (A,B) model, A describing the $z^{-3}$ dependence for the vW shift, and B the $z^{-3}$ surface-induced energy transfer (affecting the optical width). Various fittings appear to be acceptable: (a) for A= 25.5, choosing B = 0 (*i.e.* the 1-parameter model) or B = 2, B = 8, B = 16, allow optimal lineshapes which are hardly distinguishable; nevertheless, the values for γ and for the amplitude coefficient $\mathcal{C}_1$ [and also $\mathcal{C}_1/\gamma$] vary sensitively with B; (b) for a sensitively different value of A (*i.e.* A = 20), large values for B (B = 40 = 2A) are acceptable, but B = 0 is not acceptable, and amplitudes strongly differ with the values found for A = 25.5. The residual pressure shift δremains nearly unchanged under these attempts.

dimensionless curve. This effect of the thermal transfer channel on the FM-SR amplitude justifies the significance of the amplitude comparison of the 459 nm and 456 nm lines (see *Letter* —fig. 3).

### 5) From FM-SR spectra to the extrapolation of vW measurements

#### a) *Accuracy of the fittings and of the resulting $C_3$ estimates*

Figure S-6 illustrates the typical evolution of the FM-SR spectra when the sapphire window temperature is raised from 300°C to 800°C (with the Cs reservoir kept at constant temperature). Remarkably, all hyperfine components (fully resolved for $6S_{1/2}\{F=3,4\} \to 7P_{1/2}\{F'=3,4\}$, well resolved for $6S_{1/2}\{F=3,4\} \to 7P_{3/2}\{F'=2,3,4,5\}$), behave identically, and their relative amplitudes fully agree with the theoretical predictions.

In all of our conditions, the high quality of the $C_3$ fitting is very comparable to the one previously demonstrated (*e.g.* [12,13]). Figure S-9 shows how, for a given experimental spectrum, we proceed to evaluate A, and ultimately $C_3$, through attempting several values for A. The quality of the fitting is highly sensitive to the choice of A, and one can even note that the correlated evaluation of $\gamma$ restricts even more narrowly the range of $C_3$ evaluation.

#### b) *Comparison between FM-SR amplitudes on the $6P_{1/2} \to \{7P_{1/2}$ -$7P_{3/2}\}$ doublet*

For a relevant comparison between the two fine-structure components $6P_{1/2} \to 7P_{1/2}$ and $6P_{1/2} \to 7P_{3/2}$, excited with two analogous, but different lasers ($\lambda_{opt} = 459$ nm and $\lambda_{opt} = 456$ nm), the two beams follow strictly the same path and shaping through apertures (see fig. S-3), with the FM-SR signal amplitude normalized by the amount of non resonant reflection at the window.

To compare the "normalized" amplitudes of the 459 nm and 456 nm lines (see fig. 3 in *Letter*), one needs to take into account the strengh of the appropriate (squared) dipole couplings. The (demodulated) FM-SR signal also requires normalization with respect to the amplitude of the applied FM, driven separately for the 2 lasers: for the precise measurements of these amplitudes, an auxiliary experiment of FM transmission was implemented through the ultra-stable Fabry-Perot for both lasers (whose finesse was measured both at 459 m and 456 nm).

Aside from these linear factors, the observed FM-SR lineshapes undergo distortion, which modifies shape *and* amplitudes modified, as a result of the vW interaction [22]. These changes can be accounted for by removing the respective surface interactions (which strongly differ for 459 nm and 456 nm lines), assuming that the surface interaction is solely given by the van der Waals CP interaction (*i.e.* neglecting the effect on the lifetime –eq. S.2). These vW-free amplitudes are provided by the $\mathcal{C}_1$ coeffcients –see *e.g.* figs S-6, S-8, S-9–, as defined for B = 0. This converts, for

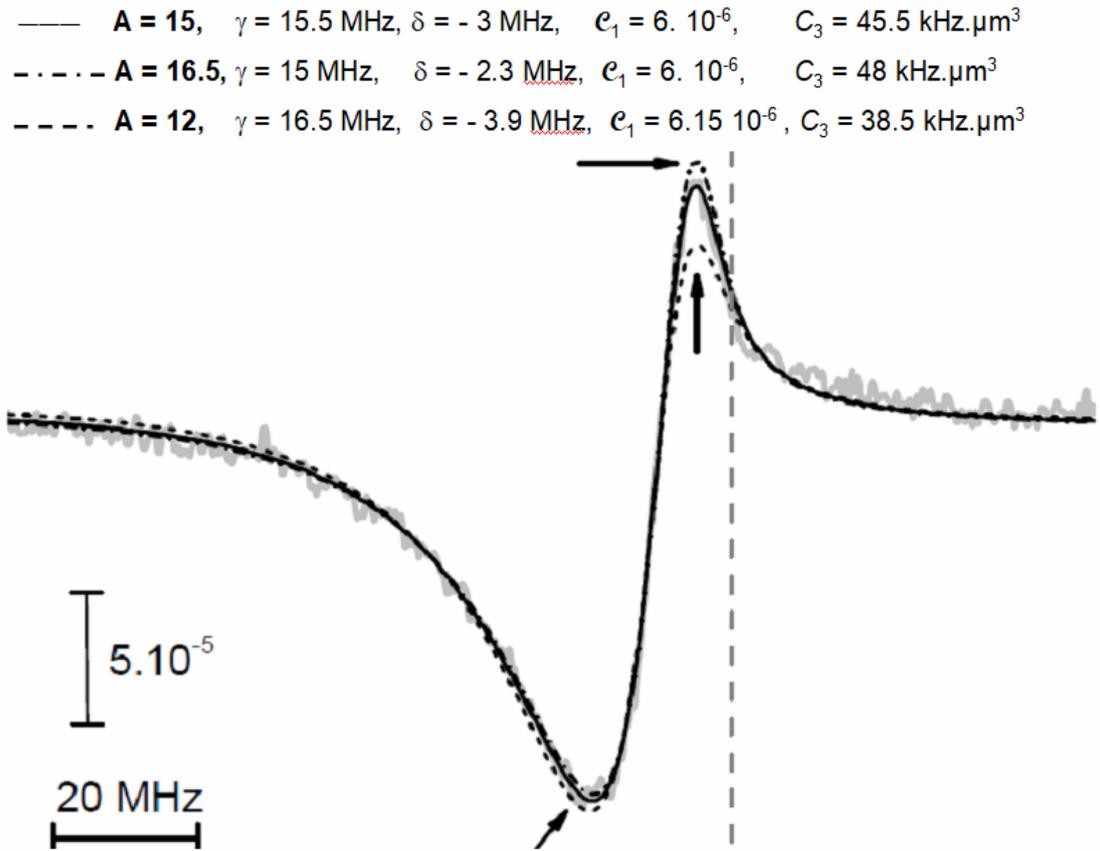

Fig. S-9: Accuracy in the fitting of a FM-SR lineshape with $- C_3 \cdot z^{-3}$ potential. The fitted experimental FM-SR spectrum is for the $6S_{1/2}$ (F=4) $\rightarrow$ $7P_{3/2}$ (F'=3) transition, at T = 300 °C for the window, and 5.3 Pa Cs pressure. The dimensionless coefficient A= 15 provides an excellent fitting, while the fittings for the neighboring values A= 16.5 and A=12, individually optimized for $\delta$, $\gamma$, $c_1$, are not satisfactory. Note that the variations over the $\gamma$ parameter cover a very moderate range, and that the resulting $C_3$ variations are smaller than those of the original dimensionless parameter A. At a given Cs density, the fitting accuracy is nearly independent of the hyperfine components, and of the surface temperature.

purpose of amplitude comparison, the actual FM-SR lineshapes into A = 0 lineshapes, which are simple dispersive (Doppler-free) Lorentzians, with respective widths $\gamma_{459}$ and $\gamma_{456}$. These widths, resulting from the lineshape fitting, are very similar at low (window) temperatures. They grow parallel with the Cs density (as governed by the reservoir temperature), and differ only slightly ($\leq 10$ %) at high window temperatures,), in agreement with the pressure broadenings found for auxiliary SA experiments (performed at rather low temperatures, under a genuine vapor equilibrium). However, these slightly distinct widths must be considered when comparing the amplitudes, because the dispersive A = 0 lineshape has an amplitude proportional to $\gamma^{-1}$; the very small difference (< 1%) between the wavevectors can be formally included, following [22], while the thermal velocity, although dependent on temperature, is identical for the two experiments when performed simultaneouly. Hence, the normalized amplitude ratio (close to unity) appearing in fig.3 (*Letter*), is provided by:

$$(\mathcal{C}_1^{459}/\mathcal{C}_1^{456}).[D_{459}(F{\to}F')/(D_{456}(F''{\to}F'''))]^{-2}.(M_{459}/M_{456})^{-1}.(\gamma_{459}/\gamma_{456}).(k_{459}/k_{456}),$$

with $M_{459}$ and $M_{456}$ the respective amplitudes of the applied FM. For this normalization factor, it is assumed that the number of active atoms is the same for both transitions. This actually implies a fully linear regime for FM-SR linearity, which is legitimated as the hyperfine optical pumping is checked to be fully negligible in our conditions (fig. S-5)

Note that in the principle, the FM-SR amplitude in the absence of surface interaction can be straightforwardly derived from the linear absorption, without a comparison to a reference line (here, the 456 nm line, whose lifetime is expected not to undergo spatial dependence). This absolute determination has to assume similar experimental conditions, and a genuine linear regime, for linear absorption and for FM-SR. The effectively recorded FM-SR spectrum should hence reveal, through its distortion and amplitude, the appropriate set of (A, B) values. This task of an *a priori* SR amplitude evaluation is actually extremely challenging, as due to various calibration issues: one difficulty is the very large difference in signal sizes, as absorption covers a macroscopic length, while reflection is only sensitive to a fraction of a wavelength. Also, this assumes the optical width $\gamma$ to remain the same in the volume and in the vicinity with the surface. This seems experimentally not always valid (see below, and the last section of [31]).

### c) Systematic uncertainties in FM-SR and how it may affect $C_3$ evaluation

FM-SR has enabled numerous experimental evaluations of the $C_3$ coefficient, in an acceptable agreement with the theoretical values (see *e.g.* [7,12,13,19,23,24]). However, consideration of systematic uncertainties has remained limited. The single observation of a major disagreement [26] was related to a notable degradation affecting the CaF$_2$ window (surface quality, and also homogeneity of the bulk), which has degraded the expected surface polariton.

We have shown above how the uncertainty on $C_3$ is determined by our fitting model (*e.g.* fig. S-9). A systematic difficulty, when converting the fitting A value into a $C_3$ coefficient, is that this demands not only reliability of the $z^{-3}$ potential (see subsection II.4 e), but also a good knowledge of $\gamma$ (see eq. S.8). In nearly all FM-SR experiments performed at low pressure, the $\gamma$ values exceed the natural atomic width, and even the expected pressure broadening [32].

At low Cs density, and for sapphire window temperature as low as ~100-150°C, the minimal FM-SR width $\gamma$, derived from our vW analysis, has always been $\geq 8$ MHz, while SA signals in a very low Cs cell temperature (30-50°C), are as narrow as 1.5 MHz (theoretical natural width ~ 1.0 MHz for the 459 nm line, ~ 1.2 MHz for 456 nm line [20]). This difference between FM-SR and SA probably corresponds to residual cell impurities, whose effect may strongly increase with temperature, rather than to the negligible self-broadening of Cs vapor (NB: the SA signal is narrow, so that the laser jitter and the FM amplitude cannot explain the observed broadening).

For a more in-depth investigation of issues related to the measured width in FM-SR, we have recently performed a FM-SR experiment and a simultaneous (volume) SA experiment, in the very same experimental conditions, on the presently used Cs cell [31]. This auxiliary experiment was performed on the strong $D_1$ resonance line of Cs (894 nm), at a very low reservoir temperature ~50°C (*i.e.* Cs density about 3 orders of magnitude lower than for our "low pressure" experiment on the second resonance line 6S-7P). To our surprise, we observe $\gamma_{FM-SR} > \gamma_{SA}$ (*i.e.* 8 MHz in FM-SR *vs.* 6 MHz in SA). A simple possibility would be that the vicinity with the surface increases the local density of perturbers (*e.g.* impurities desorbing from the surface), as compared to the region (few mm-long) probed in the comparable SA experiment. Such a local effect, if homogeneous over more than a few optical wavelengths, should not affect the $C_3$ estimate. Nevertheless, a key assumption for the detailed FM-SR model [22] is that the atomic transition is governed by an optical width supposed to be homogeneous over the whole velocity distribution. Actually, it cannot be ascertained that atoms with a velocity parallel to the window do not have a different chance to collide with impurities desorbing from the window, than atoms with a random velocity. Hence, despite accurate and consistent evaluations of the dimensionless parameter A, an unexplained extra-broadening in FM-SR, which exceeds the width of the atomic transition in an equivalent environment, may lead to an overestimate of the $C_3$ value.

Finally, despite the possibility of a systematic overestimate of $C_3$, the estimated values for $\gamma$ derived from FM-SR fittings remain very comparable, in a given experiment, for the two components of the fine-structure doublet $6S_{1/2} \rightarrow \{7P_{1/2}, 7P_{3/2}\}$. Hence, aside from a possible recalibration in the $C_3$ scale (as it may occur for the T scale –section II.1), this should not affect our conclusions concerning the relative amplitudes of the doublet components.

**III. From emissivity measurements to temperature-dependent surface response of sapphire**

1) <u>Sapphire samples and the spectrometer</u>

To minimize systematic uncertainties induced by the specificities of a given sapphire sample, the sapphire thermal emissivity was measured with a state-of-the-art Fourier-transform spectrometer [33]. The set-up presently used is improved relatively to the one used in [14]. It notably allows high-temperature studies through heating by a $CO_2$ laser, and provides a measurement of the thermal emissivity spectrum, rather than of reflectance [14] —leading to noisy spectra only at low temperatures. Although it can cover a broad spectral range, measurements are practically limited to the 5-25 µm range (12-60 THz): sapphire transparency below 5 µm ( >60 THz) makes emissivity not measurable; also for instrumental reasons, large spurious oscillations appear below 12 THz ( >25 µm), that are due to the presence of KBr optical windows along the optical path of the whole set-up, and whose effects become major when the emissivity signal is low. As we are mostly interested in the surface response for atomic transitions at 20-30 THz, this limitation could seem minor at the times of these measurements, but may have a stronger than expected influence for the extraordinary permittivity $\varepsilon_e$ (see *Letter*).

Measurements were performed on a spare window ($c_\perp$) for the Cs cell, which is strictly identical to the one used for the Cs cell. Remaining differences between this sapphire sample, and the window ending our Cs cell, may lie in the stress (mechanical or thermal) induced by the gluing of the window to the tube —possibly allowing for extrinsic resonances [10]—, or in a possible chemical transformation induced by the Cs vapor environment. The elevated temperature of the window, with respect to remote parts of the cell body and Cs reservoir, tends nevertheless to favor desorption of the impurities, which could accumulate elsewhere in colder regions of the cell.

2) <u>Sapphire birefringence</u>

Sapphire is birefringent, but in the visible range, *ordinary* and *extraordinary* indices exhibit only very small differences (relative index difference $\sim 5.10^{-3}$). Oppositely, at low frequencies including the mid- or far infrared range, differences are considerable, with a $\sim 20$ % difference for static permittivity. Owing to differences in symmetries, the *ordinary* (relative) permittivity ($\varepsilon_o$) involves 4 intrinsic modes for the bulk resonances, *vs.* only 2 for the *extraordinary* one ($\varepsilon_e$) [10].

For a window whose *c*-axis is strictly perpendicular to the surface, the cylindrical symmetry of the vW interaction is not affected by the birefringence, and an effective permittivity $\varepsilon_{eff} = (\varepsilon_o . \varepsilon_e)^{1/2}$ allows generalizing the results for an isotropic interface [24,34]. Note that if the ideal cylindrical symmetry of the Casimir-Polder atom-surface interaction is broken, the atom-surface interaction becomes more complex. However, for a spherical atom (*i.e.* S state), an "effective permittivity" remains, defined by a complex weighting of $\varepsilon_o$ and $\varepsilon_e$ [24]. In the case of our Cs cell window, the *c*-axis is perpendicular to the surface with accuracy better than 1°.

The (normal incidence) emissivity spectra of the $c_\perp$ window yields information related to $\varepsilon_o$ only. For the $c_{//}$ window, the emissivity mixes contributions associated to $\varepsilon_o$ and $\varepsilon_e$, which can be discriminated through polarization: the spectrum of emissivity observed through a polarizer parallel to the *c*-axis yields an information related to $\varepsilon_e(\omega)$, and the information related to $\varepsilon_o(\omega)$ is obtained with a polarizer perpendicular to the *c*-axis. We have verified that the emissivity for the $c_{//}$ window, as transmitted by the polarizer yielding $\varepsilon_o$, was fully similar to the one with the $c_\perp$ window; we have essentially used the initial information, from the spare Cs cell window ($c_\perp$), for information related to $\varepsilon_o(\omega)$.

Figure S-10 presents a full set of temperature evolution for emissivity spectra (fig.S-10 a), along with complex relative dielectric permittivity (fig.S-10 b), and figure S-11 shows the calculated surface responses. These two figures extend what is exemplified in the *Letter* (fig. 2), for two (approximate) temperatures ~ 500 K and ~ 1000 K. Because the laser heating of the window implies a temporal evolution, and a transient thermal equilibrium (for which temperature can be measured with accuracy $\leq$ 5K), spectra yielding $\varepsilon_o$ and $\varepsilon_e$ could not be recorded at identical temperatures. Hence, when evaluating $\varepsilon_{eff}$, we use $\varepsilon_{eff}(T) = [\varepsilon_o(T_1) . \varepsilon_e(T_2)]_e^{1/2}$ with $T_1 \sim T_2$, the indicated temperature T being an average of $T_1$ and $T_2$ [see Fig. S-11 b, and also fig.2 (*Letter*)]. This is why, in fig.2 (*Letter*), the temperatures, roughly indicated to be ~ 500 K and ~ 1000 K, correspond respectively to 490K and 949 K for ordinary axis, and to 550 K and 994 K for extraordinary axis (*i.e.* ~ 520 K and ~ 970 K for the surface response derived from $\varepsilon_{eff}$).

### 3) From emissivity measurements to the relative dielectric permittivity $\varepsilon(\omega)$

The complex value of the relative permittivity $\varepsilon(\omega)$ can be extrapolated from the fitting of an analytical function, able to describe the reflectance spectrum $R(\omega)$, or the emissivity spectrum $E(\omega) = 1 - R(\omega)$, through an appropriate function for $\varepsilon(\omega)$. Our method has been described previously, with the widths of the bulk resonances allowing for thermal phonon contributions through a Kramers-Kronig Gaussian distribution (see [14] and refs. therein). This approach has recently proved valuable in a comparative study [35].

To describe the relative dielectric permittivity, the resonances associated to the intrinsic bulk modes for the IR range (4 or 2, respectively for ordinary and extraordinary sapphire), are summed-up, in the frame of a modified Lorentz dielectric model including Gaussian multiphonon contribution *via* self-energy-function (ee *e.g.* [14]):

$$\varepsilon(\omega) = \varepsilon_\infty + \sum_j \frac{S_j \Omega_j^2}{\Omega_j^{\,2} - \omega^2 - 2\Omega_j P_j(\omega)} \qquad (S.10)$$

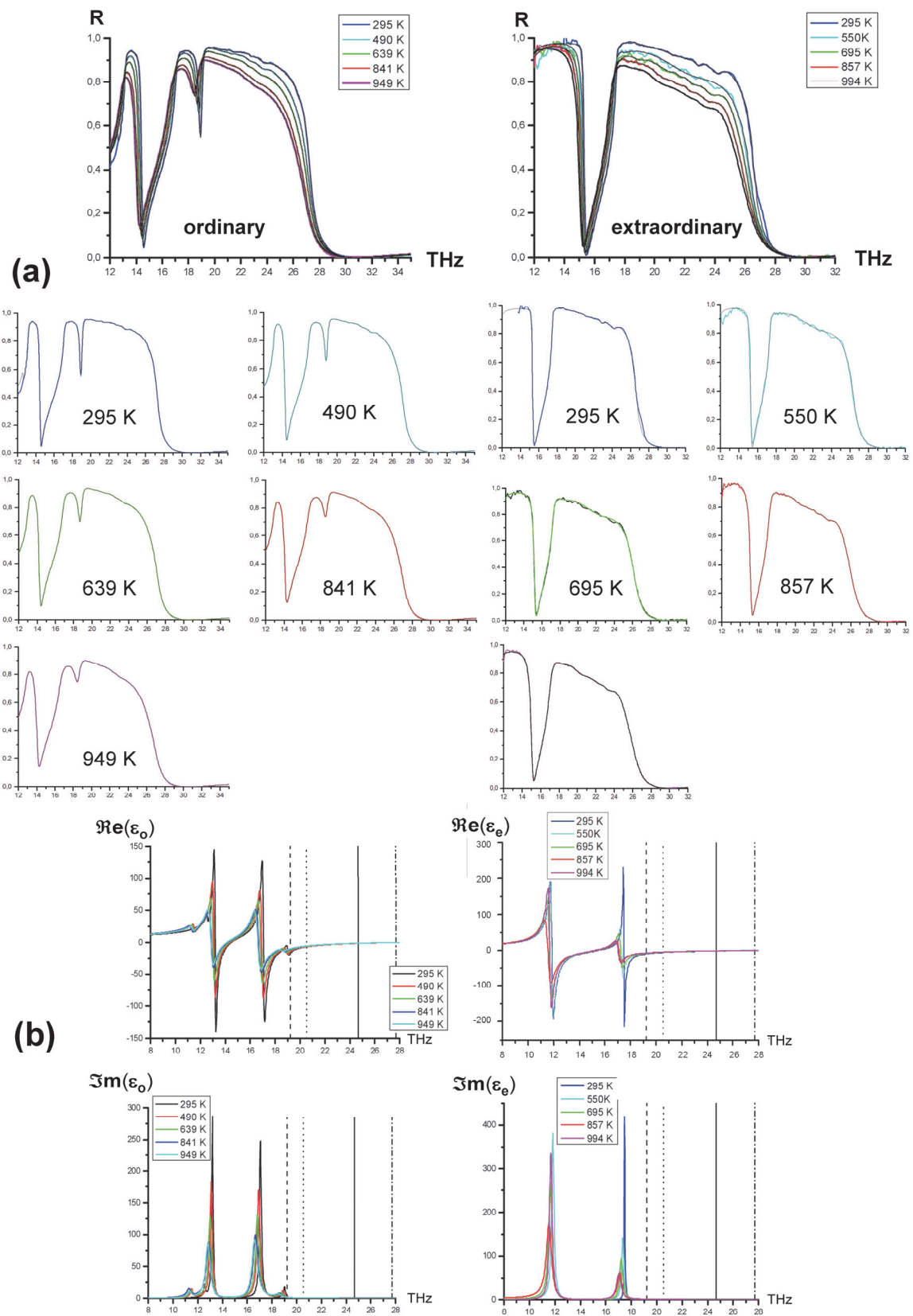

Fig. S-10: (a) Evolution of the Reflectance spectrum with temper a t u r e ( a s  i n d i c a t e d ) ; a l l  e x p e r i m e n t a l  spectra (in color) are fitted with a black line, hard-to-see because of the quality of the fits; the spectra are also presented individually to better visualize the quality of these fits: (left) reflectance from the $c_\perp$ window, *i.e.* ordinary axis; (right) from the $c_{//}$ window and a polarizer selecting the information relevant for the extraordinary axis; (b) Relative dielectric (complex) permittivity spectra, as derived from the fits for emissivity; the temperature evolution for $\Re[\omega(\varepsilon)]$ and $\Im[\omega(\varepsilon)]$ is shown on the left for $\varepsilon_o$ (described by a 4-phonon model), on the right for $\varepsilon_e$ (described by a 2-phonon model). The vertical lines mark the atomic transitions of interest: full line at 24.687 THz for the $7P_{1/2} \rightarrow 6D_{3/2}$ transition; dashed one and dotted one, respectively at 19.24 THz and 20.54 THz, for the $7P_{3/2} \rightarrow \{6D_{3/2}, 6D_{5/2}\}$ couplings; dashed-dotted line at 27.68 THz for the $7D_{3/2} \rightarrow 5F_{5/2}$ coupling, analyzed in [12].

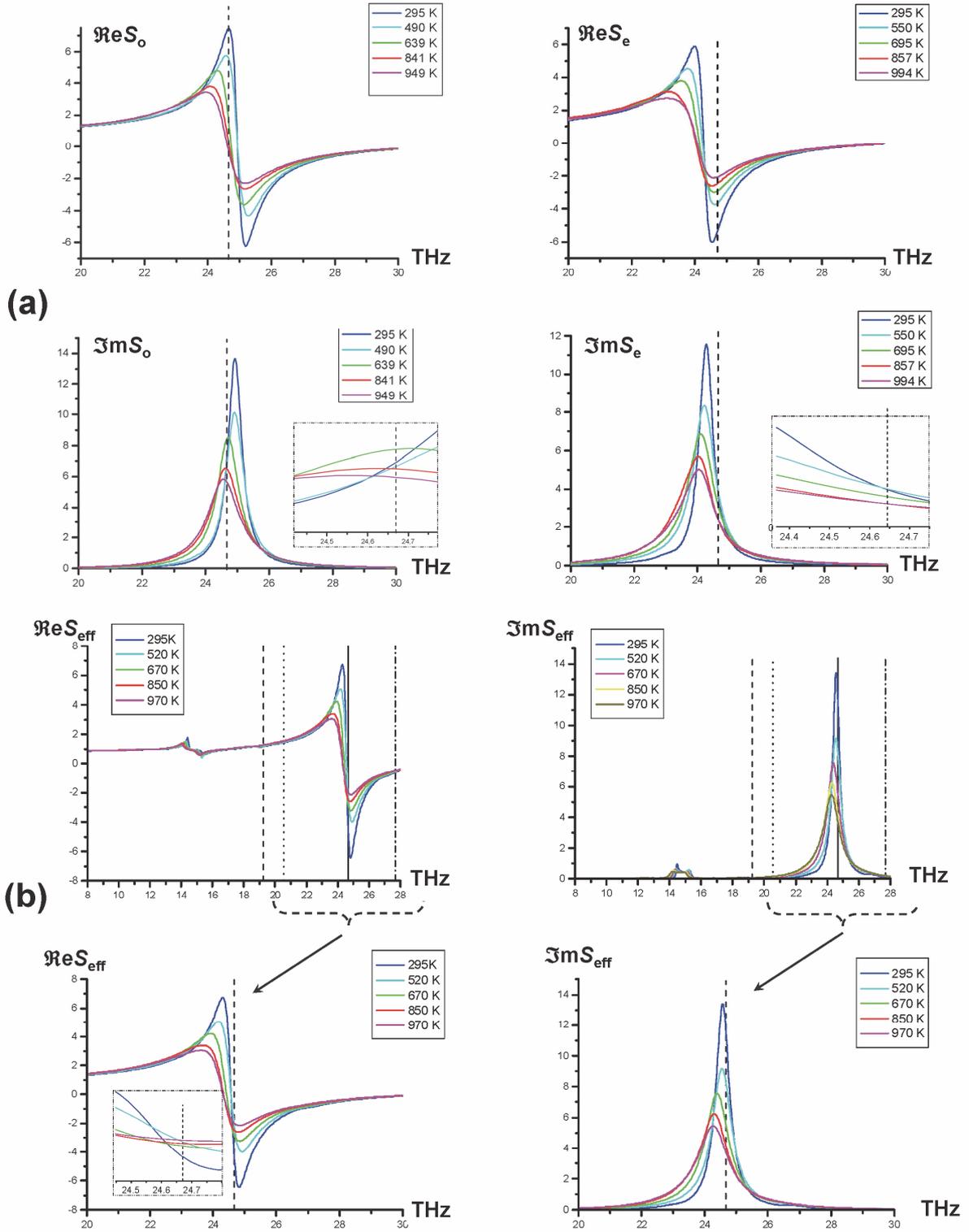

Fig. S-11: Temperature evolution of the surface response $S(\omega) = [\varepsilon(\omega) - 1] / [\varepsilon(\omega) + 1]$, as calculated from the dielectric permittivity spectra of fig. S-10. (a) Complex surface response calculated for a fictitious isotropic medium, defined either by $\varepsilon_o$ (left), or by $\varepsilon_e$ (right); (b) Surface response for a $c_\perp$ window, with $S_{\text{eff}}(\omega)$ defined by the effective relative permittivity $\varepsilon_{\text{eff}} = (\varepsilon_o.\varepsilon_e)^{1/2}$. On top, the extended range of the spectrum allows to visualize small and distorted resonances in the wings of the principal resonance; on the bottom, the zoom for $\Re[S(\omega)]$ around the major resonance illustrates how the simultaneous temperature shift and broadening of the resonance can lead to non monotonic temperature variations. The frequency markers are the same as in fig. S-10.

In eq. (S.10), $\varepsilon_\infty$ is the high-frequency value of the dielectric constant, $\Omega_j$ the transverse optical frequency of the bulk resonance, $S_j$ the dielectric strength, and $P_j(\omega)$ the self-energy of each resonance, which generalizes a damping parameter. $P_j(\omega)$ involves several multi-phonon contributions, whose importance grows up with temperature, and consists of a sequence of peaks that are well reproduced by "extended" Gaussian functions:

$$P_j(\omega) = \sum_n \widetilde{g}_{n,j}(\omega) \qquad (S.11)$$

with:

$$\widetilde{g}_{n,j}(\omega) = \frac{2A_{n,j}}{\sqrt{\pi}} \left[ D\left( \frac{2\sqrt{Ln(2)}(\omega + \omega_0^{n,j})}{\gamma_{n,j}} \right) - D\left( \frac{2\sqrt{Ln(2)}(\omega - \omega_0^{n,j})}{\gamma_{n,j}} \right) \right]$$
$$+ A_{n,j}\, i \exp\left( -\frac{4Ln(2)(\omega - \omega_0^{n,j})^2}{\gamma_{n,j}^2} \right) - A_{n,j}\, i \exp\left( -\frac{4Ln(2)(\omega + \omega_0^{n,j})^2}{\gamma_{n,j}^2} \right) \qquad (S.12)$$

In eq.(S.12), $A_{n,j}$ is the amplitude, $\omega_0^{n,j}$ the location, $\gamma_{n,j}$ the full width at half-maximum parameters of Gaussians, with $D(x)$ the Dawson integral:

$$D(x) = \exp(-x^2)\int_0^x \exp(t^2)dt \qquad (S.13)$$

### 4) From permittivity $\varepsilon(\omega)$ to the surface response $S(\omega)$

In the principle, with an analytical function describing the (relative) permittivity $\varepsilon(\omega)$, the surface response $S(\omega)$ is obtained straightforwardly. Nevertheless, the key features for $S(\omega)$ appear away from resonances for $\varepsilon(\omega)$. It is hence important to investigate how subtle choices for the modeling of $\varepsilon(\omega)$ affect predictions for the surface response. An *a priori* estimate of the sensitivity to these changes can hardly be given, and another difficulty is that, with respect to the large number of fitting parameters, the uncertainty can be important for some of the parameters.

Practically, various sets of parameters allow good quality fittings for the emissivity. These differing sets affect weakly the amplitude and widths of resonances for $S(\omega)$. This insensitivity of the surface resonance may originate in the fact that reflectance (or emissivity) is already a kind of surface measurement, rather than a genuine bulk information: it is measured under the normal incidence, while the surface response $S(\omega)$ corresponds to a summing of emission over all incidences —notably including incidences above the critical angle, allowing the contribution of evanescent waves.

Although the overall shape for $S(\omega)$ remains rather stable when exploring different sets of fitting parameters, the position of the surface resonance can be slightly shifted, depending on the choice of fitting parameters. The uncertainty may become comparable to the width of the surface resonance. This becomes a highly sensitive issue when predicting the surface response at a single arbitrary frequency $\omega_{at}$, as the $7P_{1/2}$-$6D_{3/2}$ resonant coupling. General considerations [36] show that $S(\omega)$ is nearly a complex Lorentzian around resonance, provided that the resonance is sharp enough —a situation which applies to sapphire, even at high-T. Close to resonance, this implies a tight connection between $\Re e S(\omega)$ and $\Im m S(\omega)$, at the origin of our suspicion (see *Letter*) of a distance larger than expected distance, between the sapphire resonance peak, and the $7P_{1/2}$-$6D_{3/2}$ transition.

As a general trend, we observe on fig. S-11 that $S(\omega)$ broadens with temperature, and also shifts towards the lower frequencies when increasing temperature. Because the atomic transition of interest for $7P_{1/2}$ is located in the negative region of the anomalous dispersion for $\Re e S(\omega)$, the decrease of $\Re e S(\omega)$ amplitude with temperature is susceptible to compete with the overall frequency shift of the resonance. Note that for a limited temperature range, this competition allows $|\Re e S(\omega_{at})|$ to increase with temperature (see zooms in fig. S-11).

## 5) Uncertainty issues and comparison with literature

### a) Experimental uncertainties affecting the emissivity spectrometer, and the related permittivity

The emissivity spectrometer, described in detail in [33], is a Fourier-transform instrument covering a broad spectral range thanks to different detectors. The spectrometer is made free of systematic frequency errors with a He-Ne laser calibration. The spectral resolution has been chosen to be 4 cm$^{-1}$ (0.12 THz) because the duration of recordings are limited by the thermal evolution of the heated sample. This limited resolution can induce a broadening for some of the resonances appearing in the dielectric function, justifying some uncertainty, but this is effective only at moderate temperatures. Nevertheless, the unambiguous temperature broadening of the extrapolated response $S(\omega,T)$ (see fig. S-10) justifies that our frequency resolution is sufficient.

Another major uncertainty lies in the calibration of amplitude. Spectral emissivity is indeed measured through a comparison (accuracy estimated to be ~1%) with "ideal blackbody emission", as generated with an auxiliary thermal source of "blackbody radiation" [33,37]. Over the large spectral range which is covered, the linear sensitivity of detectors may be imperfect. We have attempted a fitting of spectra with modified amplitude, introducing artificial changes ($\leq 1\%$) in the relative amplitude, or in the absolute one through an offset. Such artificial changes do not substantially modify the properties of $S(\omega)$, but for a possible slight shift (*i.e.* shape and amplitude unchanged). A limit of these artificial changes is that they were applied only in a homogeneous manner over the whole spectrum, discarding the spectral variations of the detector.

The angular acceptance of the spectrometer is another source of experimental uncertainty for the spectrum itself. The analysis of emissivity relies on the collection of the field (thermally) emitted under near normal incidence. The acceptance angle is actually ~ 4° (half-angle). A rather large acceptance angle (~ 11°) has been shown [35] to have a negligible influence, at least for isotropic materials ($CaF_2$ and $BaF_2$). For sapphire, the spectra for $\varepsilon_o$ and $\varepsilon_e$ differ strongly:

collecting the emissivity at non-normal incidence —or with low polarization selectivity— may induce a mixing of these very different spectra associated to orthogonal orientations. We believe that this contribution to uncertainty is rather low for our conditions: for $\varepsilon_o(\omega)$, we have found a sound agreement between the 2 different measurements ($c_\perp$ sample, and $c_{//}$ sample with the adequate polarizer); for $\varepsilon_e(\omega)$, the possible contamination by emissivity associated to $\varepsilon_o(\omega)$ cannot be evaluated by a comparison, but emissivity spectra for the 4-phonon model exhibit typical features, notably around 18-20 THz, which are clearly absent on the spectra yielding $\varepsilon_e(\omega)$ —see fig. S-10.

As mentioned in the *Letter*, fig. S-10 shows that the major contribution to $\varepsilon_e(\omega)$ comes from a resonance at $\leq 12$ THz, notably when the temperature is high, while the secondary peak (around 18 THz) severely decreases with temperature. This situation, not previously reported or anticipated, differs from the parallel temperature evolution exhibited by the peaks of $\varepsilon_o(\omega)$, at 13 THz and 17 THz (see *e.g.* [14]). It shows that the evaluation of $\varepsilon_e(\omega)$ is actually derived from an extrapolation to a frequency domain where measurements were actually not performed, and this can add notably to the uncertainty issues. In this sense, one can even note that the dispersion-like peak around 12 THz (fig. S-10) is exceptionally low for 857 K, while its amplitude remains nearly constant for other temperatures. This may reveal some inconsistency in the choice of the multiple fitting parameters, but an analogous situation was already found for 626 K (in comparison with measurements at 472K and 870 K) in the (unpublished) course of the work leading to [37] (measurements performed at lower temperatures than those of interest for [37]). The absence of reliable emissivity measurements below ~12 THz could have a role in this double occurrence of a slight inconsistency.

*b) Systematic attempts to modify the fitting values*

Although the emissivity spectrometer is broadband and despite the thermal emission is effective only in a limited energy range, the Kramers-Kronig relationship demands integration over the whole spectrum. The UV absorption, which is nearly independent of temperature, and the dispersion in the transparency range ($\geq 60$ THz) have not been included in our approach. Rather, only a single parameter is considered for the asymptotic blue-wing behavior, through $\varepsilon_\infty$ [see eq. (S.10)]. Following the same vein of arbitrary changes as reported in the previous subsection for the experimental emissivity and its amplitude, we have attempted to modify $C_3(T)$ predictions by allowing an arbitrary change in $\varepsilon_\infty$ (see eq. S.10), without a significant success. The major limit of such arbitrary changes is that, away from the IR range, the properties of sapphire are actually constrained on the one hand by the static permittivity, on the other hand by the slowly dispersive refractive index for the visible range and the UV absorption [38]. These quantities can be refractive index for the visible range and the UV absorption [38]. These quantities can be determined through independent measurements, and should not sensitively vary with temperature in the UV range.

### c) Other sources from literature

Before the present temperature study for sapphire, some of us had already investigated, with a reflectance spectrometer, the temperature behavior of sapphire, using the same model (eq. S.10) for dielectric permittivity. The initial study [14] was limited to T ≤ 500°C, for a $c_\perp$ window of an uncharacterized origin. A further study [37], using the same spectrometer as now used, provided novel results for ordinary and extraordinary orientations, in view of an improved understanding of processes leading to melting at very high temperatures (≥ 2000 K); published data had remained limited for temperatures below 1200 K. Our present set of data, with more temperatures below 1200 K, and for windows from the same origin as the one ending the Cs cell, is in overall agreement with the (*unpublished*) set of parameters fitting these emissivity measurements (see fig. S-12).

Before our works, the temperature dependence of the optical properties for sapphire was also investigated systematically in several works [39-41]:

- Emissivity at several high temperatures was notably measured, limited to an ordinary ray sapphire sample [39]. The accuracy of the fittings for emissivity is not as satisfactory as those shown in fig. S-10; fittings in [39] involve a spectral temperature dependence for the index and absorption coefficient of the material –without a direct attempt to deal with the dielectric permittivity.

- A complementary analysis of the temperature-dependence of the refractive index measurements, limited to the transparency range of sapphire (*i.e.* visible and near IR), for a sapphire sample of unknown origin, was presented in [40]. The analysis is based on a classical oscillator model, not as elaborated for the IR range as our Kramers-Kronig Gaussian expansion. Importantly, it has the advantage to propose a temperature expansion of parameters, up to the second order, for the various coefficients affecting each of the resonances, up to the electronic absorption (UV range). It is supposed to describe the entire spectrum for the permittivity, as in a generalized Sellmeier approach. This model, solely based upon measurements in the transmission region, includes the two UV resonances, but allows 5 and 3 modes in the IR range (respectively for $\varepsilon_o$ and $\varepsilon_e$), despite the seminal analysis of [10] allowing only respectively 4 and 2 intrinsic resonances for sapphire. Because of the convenience of this model, which provides a continuous evolution of the surface response with temperature, we compare (see fig. S-12) the temperature-dependent surface response $S(\omega_{at})$ at the specific frequency $\omega_{at}$ of the resonant coupling $7P_{1/2} \rightarrow 6D_{3/2}$ (24.687 THz), as derived from [40], with the results of our own analysis. Despite important variations, the two predictions seem to get closer for those high temperatures, when the "giant" $C_3(T)$ coefficient is truly sensitive to the temperature-broadened sapphire resonance (see fig.1 of *Letter*).

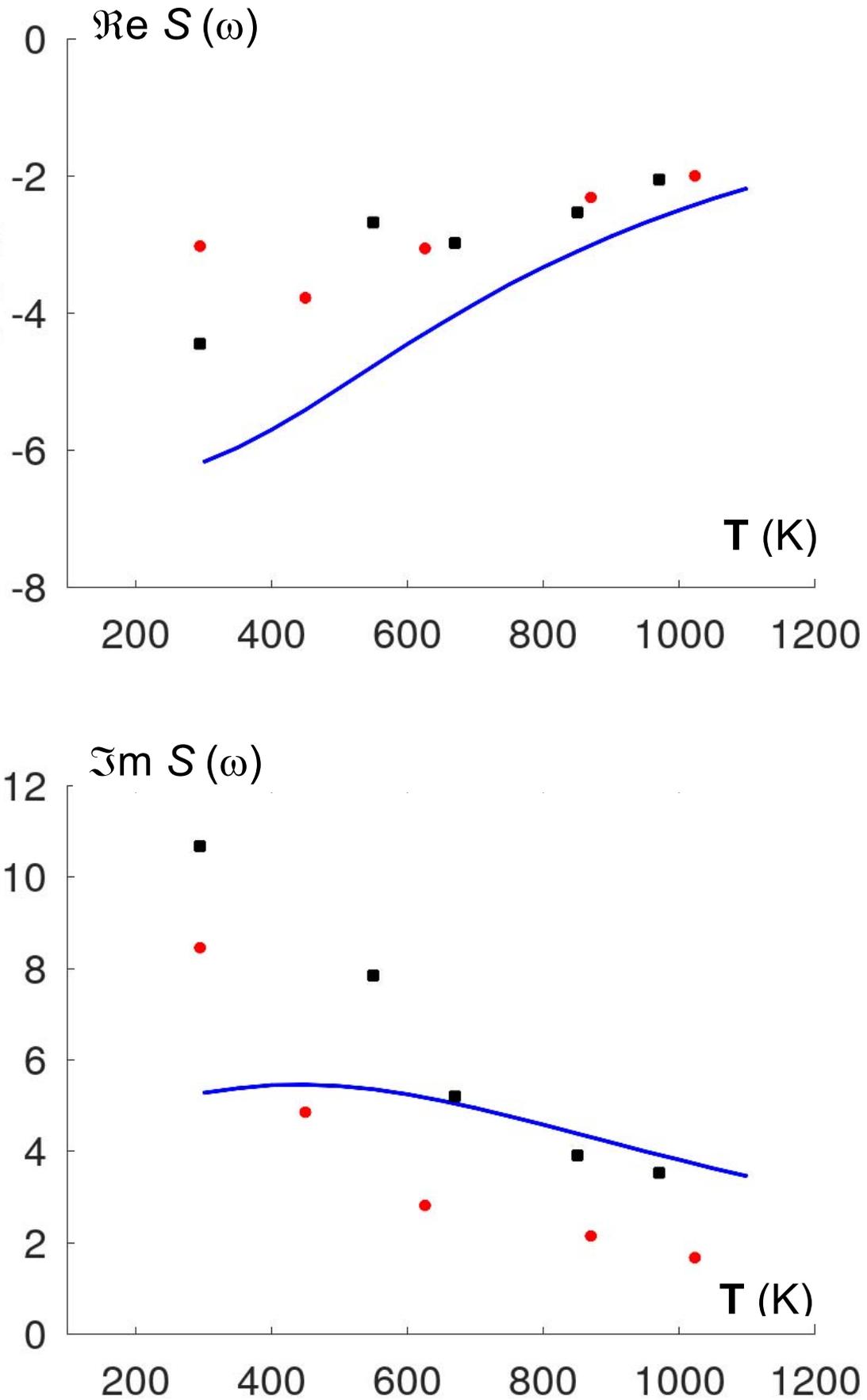

Fig. S-12 $\Re eS(\omega_{at},T)$ and $\Im mS(\omega_{at},T)$, as a function of temperature for $c_\perp$-axis sapphire [*i.e.* $\varepsilon_{eff} = (\varepsilon_o \cdot \varepsilon_e)^{1/2}$] at $\omega_{at} \approx 24.687$ THz (*i.e.* Cs coupling $7P_{1/2} \rightarrow 6D_{3/2}$); the continuous blue line uses $\varepsilon_o(T)$ and $\varepsilon_e(T)$ as derived from [40], the red points are from our present measurements, and the black squares from measurements acquired at the time of [37].

- At last, it is worth mentioning that in a rather recent work [41], a Lorentz model is used to combine previously obtained UV data, with data derived from ellipsometry, for experiments performed up to 573 K. This temperature range is too limited to be worth a comparison with our results. Also, the specific benefit of ellipsometry, with its ability to provide directly the complex refractive index at the frequency of interest (*i.e.* around the surface resonance) is here lost by the use of "spectroscopic ellipsometry", where the *a priori* Lorentzian model is applied to provide the permittivty ε, hence giving a considerable weight onto the bulk resonance, and not on the surface resonance at the core of our work.

resonances, down to the visible range, were already analyzed –see A. K. Harman, S. Ninomiya, S. Adachi, *Optical constants of sapphire (α-Al₂O₃) single crystals* J. Appl. Phys. **76**, 8032 (1994).